\documentclass[11pt,onecolumn,draftclsnofoot]{IEEEtran}
\usepackage{amsmath,amssymb} \interdisplaylinepenalty=2500
\usepackage{cite}
\usepackage{bm}
\usepackage{color}

\newtheorem{definition}{Definition}
\newtheorem{theorem}{Theorem}
\newtheorem{proposition}[theorem]{Proposition}
\newtheorem{corollary}[theorem]{Corollary}
\newtheorem{lemma}[theorem]{Lemma}

\newtheorem{example}{Example}
\newtheorem{remark}{Remark}
\def\qed{\endIEEEproof}

\DeclareMathOperator*{\argmin}{argmin}
\renewcommand{\dim}{\operatorname{\sf dim}\hspace{0.1em}}
\DeclareMathOperator{\rank}{\sf rank\hspace{0.1em}}
\DeclareMathOperator{\wt}{\sf wt\hspace{0.1em}}
\DeclareMathOperator{\minpoly}{\sf minpoly}
\DeclareMathOperator{\RRE}{\sf RRE}

\newcommand{\nlinks}{\ell}
\newcommand{\Fq}{\mathbb{F}_q}
\newcommand{\Fqm}{\mathbb{F}_{q^m}}
\newcommand{\linspan}[1]{\left< #1 \right>}
\newcommand{\mat}[1]{\begin{bmatrix} #1 \end{bmatrix}}
\newcommand{\PP}{\mathcal{P}}
\newcommand{\ds}{d_{\scriptscriptstyle \textrm{S}}}
\newcommand{\dr}{d_{\scriptscriptstyle \textrm{R}}}
\newcommand{\I}{\mathcal{I}}

\newcommand{\defn}{\textit}  %% use for terms defined inline
\title{A Rank-Metric Approach to Error Control in Random Network Coding}

\author{Danilo~Silva,~\IEEEmembership{Student~Member,~IEEE,}
        Frank~R.~Kschischang,~\IEEEmembership{Fellow,~IEEE,}
        and~Ralf~K{\"o}tter,~\IEEEmembership{Senior~Member,~IEEE}%
\thanks{This work was supported by CAPES Foundation, Brazil, and by the Natural Sciences and Engineering Research Council of Canada. The material in this paper was presented in part at the IEEE International Symposium on Information Theory, Nice, France, June 2007 and at the IEEE Information Theory Workshop, Bergen, Norway, July 2007.}
\thanks{D. Silva and F. R. Kschischang are with The Edward S. Rogers Sr. Department of Electrical and Computer Engineering, University of Toronto, Toronto, ON M5S 3G4, Canada (e-mail: danilo@comm.utoronto.ca, frank@comm.utoronto.ca).}
\thanks{R. K{\"o}tter is with the Institute for Communications Engineering, Technical University of Munich, D-80333 Munich, Germany (e-mail: ralf.koetter@tum.de).}
}

\begin{document}
\maketitle
\thispagestyle{empty}

\vspace{-2.5ex}
\begin{abstract}
The problem of error control in random linear network coding is addressed from a matrix perspective that is closely related to the subspace perspective of K{\"o}tter and Kschischang. A large class of constant-dimension subspace codes is investigated. It is shown that codes in this class can be easily constructed from rank-metric codes, while preserving their distance properties. Moreover, it is shown that minimum distance decoding of such subspace codes can be reformulated as a generalized decoding problem for rank-metric codes where partial information about the error is available. This partial information may be in the form of erasures (knowledge of an error location but not its value) and \emph{deviations} (knowledge of an error value but not its location). Taking erasures and deviations into account (when they occur) strictly increases the error correction capability of a code: if $\mu$ erasures and $\delta$ deviations occur, then errors of rank $t$ can always be corrected provided that $2t \leq d - 1 + \mu + \delta$, where $d$ is the minimum rank distance of the code. For Gabidulin codes, an important family of maximum rank distance codes, an efficient decoding algorithm is proposed that can properly exploit erasures and deviations. In a network coding application where $n$ packets of length $M$ over $\Fq$ are transmitted, the complexity of the decoding algorithm is given by $O(dM)$ operations in an extension field $\mathbb{F}_{q^n}$.
\end{abstract}


\section{Introduction}
\label{sec:introduction}

While random linear network coding \cite{Ho++2003,Chou++2003,Ho++2006} is an effective technique for information dissemination in communication networks, it is highly susceptible to errors. The insertion of even a single corrupt packet has the potential, when linearly combined with legitimate packets, to affect all packets gathered by an information receiver. The problem of error control in random network coding is therefore of great interest.

In this paper, we focus on end-to-end error control coding, where only the source and destination nodes apply error control techniques. Internal network nodes are assumed to be unaware of the presence of an outer code; they simply create outgoing packets as random linear combinations of incoming packets in the usual manner of random network coding.
In addition, we assume that the source and destination nodes have no knowledge---or at least make no effort to exploit knowledge---of the topology of the network or of the particular network code used in the network. This is in contrast to the pioneering approaches \cite{Cai.Yeung2002,Yeung.Cai2006,Cai.Yeung2006}, which have considered the design of a network code as part of the error control problem.

In the basic transmission model for end-to-end coding, the source node produces $n$ packets, which are length-$M$ vectors in a finite field $\Fq$, and the receiver gathers $N$ packets. Additive packet errors may occur in any of the links. The channel equation is given by $Y = AX + BZ$, where $X$, $Y$ and $Z$ are matrices whose rows represent the transmitted, received and (possibly) corrupting packets, respectively, and $A$ and $B$ are the (unknown)
corresponding transfer matrices induced by linear network coding.

There have been three previous quite different approaches to reliable communication under this model. In \cite{Zhang2008}, Zhang characterizes the error correction capability of a network code under a brute-force decoding algorithm. He shows that network codes with good error-correcting properties exist if the field size is sufficiently large. His approach can be applied to random network coding if an extended header is included in each packet in order to allow for the matrix $B$ (as well as $A$) to be estimated at a sink node. A drawback of this approach is that the extended header has size equal to the number of network edges, which may incur excessive overhead. In addition, no efficient decoding algorithm is provided for errors occurring according to an adversarial model.

Jaggi et al. \cite{Jaggi++2007:Byzantine} propose a different approach specifically targeted to combat Byzantine adversaries. They provide rate-optimal end-to-end codes that do not rely on the specific network code used and that can be decoded in polynomial time. However, their approach is based on probabilistic arguments that require both the field size and the packet length to be sufficiently large.

In contrast, K{\"o}tter and Kschischang \cite{Kotter.Kschischang2008} take a more combinatorial approach to the problem, which provides correction guarantees against adversarial errors and can be used with any given field and packet size. Their key observation is that, under the unknown linear transformation applied by random network coding, the only property of the matrix $X$ that is preserved is its row space. Thus, information should be encoded in the choice of a subspace rather than a specific matrix. The receiver observes a subspace, given by the row space of $Y$, which may be different from the transmitted space when packet errors occur. A metric is proposed to account for the discrepancy between transmitted and received spaces, and a new coding theory based on this metric is developed. In particular, nearly-optimal Reed-Solomon-like codes are proposed that can be decoded in $O(nM)$ operations in an extension field $\mathbb{F}_{q^n}$.

Although the approach in \cite{Kotter.Kschischang2008} seems to
be the appropriate abstraction of the error control problem in
random network coding, one inherent difficulty is the absence of
a natural group structure on the set of all subspaces of the ambient space $\Fq^M$. As a consequence, many of the powerful concepts of classical coding
theory such as group codes and linear codes do not naturally extend
to codes consisting of subspaces.

In this paper, we explore the close relationship between subspace
codes and codes for yet another distance measure:  the rank metric.
Codewords of a rank metric code
are $n \times m$ matrices and the rank distance between two matrices
is the rank of their difference.
The rank metric was introduced in coding theory by
Delsarte \cite{Delsarte1978:BilinearForms}. Codes for the rank metric
were largely developed by Gabidulin \cite{Gabidulin1985}
(see also \cite{Delsarte1978:BilinearForms,Roth1991:MaximumRankArrayCodes}).
An important feature of the coding theory for the
rank metric is that it supports many of the powerful concepts and techniques of classical coding theory, such as linear and cyclic codes and corresponding decoding algorithms \cite{Gabidulin1985,Roth1991:MaximumRankArrayCodes,Richter.Plass2004:BerlekampMassey,Loidreau2005}.

One main contribution of this paper is to
show that codes in the rank metric can be naturally ``lifted''
to subspace codes in such a way that the rank distance between two codewords
is reflected in the subspace distance between their lifted images.
In particular, nearly-optimal subspace codes can be obtained directly
from optimal rank-metric codes.
Conversely, when lifted rank-metric codes are used, the decoding problem for random network coding can be reformulated purely in rank-metric terms,
allowing many of the tools from the theory of rank-metric
codes to be applied to random network coding.

In this reformulation, we obtain a generalized decoding problem for rank-metric codes that involves not only ordinary rank errors,
but also two additional phenomena that we call \emph{erasures} and
\emph{deviations}.
Erasures and deviations are dual to each other and correspond to
partial information about the error matrix, akin to the role played
by symbol erasures in the Hamming metric. Here, an erasure corresponds
to the knowledge of an error location but not its value,
while a deviation correspond to the knowledge of an error value but
not its location. These concepts generalize similar concepts found in the rank-metric literature under the terminology of ``row and column erasures'' \cite{Gabidulin++1992,Gabidulin.Pilipchuk2003,Richter.Plass2004:BerlekampMassey,Richter.Plass2004:ColumnErasures, Pilipchuk.Gabidulin2007}. Although with a different terminology, the concept of a deviation (and of a code that can correct deviations) has appeared before in \cite{Roth.Seroussi1996:LocationCorrecting}.

Our second main contribution is an efficient decoding algorithm for rank-metric codes that takes into account erasures and deviations. Our algorithm is applicable to Gabidulin codes \cite{Gabidulin1985}, a class of codes, analogous to conventional Reed-Solomon codes, that attain maximum distance in the rank metric. We show that our algorithm fully exploits the correction capability of Gabidulin codes; namely, it can correct any pattern of $\epsilon$ errors, $\mu$ erasures and $\delta$ deviations provided $2\epsilon + \mu + \delta \leq d-1$, where $d$ is the minimum rank distance of the code. Moreover, the complexity of our algorithm is $O(dM)$ operations in $\mathbb{F}_{q^n}$, which is smaller than that of the algorithm in \cite{Kotter.Kschischang2008}, especially for practical high-rate codes.

In the course of setting up the problem, we also prove a result that can be seen as complementary to \cite{Kotter.Kschischang2008}; namely, we relate the performance guarantees of a subspace code with more concrete network parameters such as the maximum number of corrupting packets that can be injected in the network. This result provides a tighter connection between the subspace approach of \cite{Kotter.Kschischang2008} and previous approaches that deal with link errors.

The remainder of this paper is organized as follows. In Section \ref{sec:preliminaries}, we provide a brief review of rank-metric codes and subspace codes. In Section \ref{sec:random-network-coding}, we describe in more detail the problem of error control in random network coding, along with K{\"o}tter and Kschischang's approach to this problem. In Section \ref{sec:rank-metric-approach}, we present our code construction and show that the resulting error control problem can be replaced by a generalized decoding problem for rank-metric codes. At this point, we turn our attention entirely to rank-metric codes. The generalized decoding problem that we introduce is developed in more detail in Section \ref{sec:generalized-decoding}, wherein the concepts of erasures and deviations are described and compared to related concepts in the rank-metric literature. In Section \ref{sec:decoding-error-erasures-deviations}, we present an efficient algorithm for decoding Gabidulin codes in the presence of errors, erasures and deviations. Finally, Section \ref{sec:conclusions} contains our conclusions.

\section{Preliminaries}
\label{sec:preliminaries}

\subsection{Notation}
\label{sec:notations}

Let $q \geq 2$ be a power of a prime.
In this paper, all vectors and matrices have components in the
finite field $\Fq$, unless otherwise mentioned.
We use $\Fq^{n \times m}$ to denote the set of all $n \times m$
matrices over $\Fq$ and we set $\Fq^n = \Fq^{n \times 1}$.
In particular, $v \in \Fq^n$ is a column vector
and $v \in \Fq^{1 \times m}$ is a row vector.

If $v$ is a vector, then the symbol $v_i$ denotes the $i$th entry of $v$.
If $A$ is a matrix, then the symbol $A_i$ denotes either the $i$th row
or the $i$th column of $A$; the distinction will always be clear from the way
in which $A$ is defined.
In either case, the symbol $A_{ij}$ always refers to the entry in the
$i$th row and $j$th column of $A$.

For clarity, the $k \times k$ identity matrix is denoted by $I_{k \times k}$.
If we set $I = I_{n \times n}$, then the notation $I_i$ will denote
the $i$th column of $I$.  More generally,
if $\mathcal{U} \subseteq \{1,\ldots,n\}$,
then $I_{\mathcal{U}} = [I_i,\, i \in \mathcal{U}]$ will denote the
sub-matrix of $I$ consisting of the columns indexed by $\mathcal{U}$.

The linear span of a set of vectors $v_1,\ldots,v_k$ is
denoted by $\linspan{v_1,\ldots,v_k}$.
The row space, the rank and the number of nonzero rows of a matrix
$X$ are denoted by $\linspan{X}$, $\rank X$ and $\wt(X)$, respectively. The reduced row echelon (RRE) form of a matrix $X$ is denoted by $\RRE(X)$.

\subsection{Properties of Matrix Rank and Subspace Dimension}
\label{sec:properties}

Let $X \in \Fq^{n \times m}$.
By definition, $\rank X = \dim \linspan{X}$;  however, there are
many useful equivalent characterizations.
For example, $\rank X$ is the smallest $r$ for which there exist
matrices $A \in \Fq^{n \times r}$ and $B \in \Fq^{r \times m}$ such
that $X = AB$, i.e.,
\begin{equation}\label{eq:rank.equivalent-definition}
  \rank X = \min_{\substack{r, A \in \Fq^{n \times r}, B \in \Fq^{r \times m}: \\ X = AB}} r.
\end{equation}

It is well-known that, for any $X,Y \in \Fq^{n \times m}$, we have
\begin{equation}\label{eq:rank.triangle-inequality}
  \rank(X + Y) \leq \rank X + \rank Y
\end{equation}
and that, for $X \in \Fq^{n \times m}$ and $A \in \Fq^{N \times n}$, we have
\begin{equation}\label{eq:rank.product-inequality}
  \rank(AX) \geq \rank A + \rank X - n.
\end{equation}

Recall that if $U$ and $V$ are subspaces of some fixed vector space, then the sum
\begin{equation}\nonumber
  U + V = \{u+v \colon u \in U,\, v \in V\}
\end{equation}
is the smallest subspace that contains both $U$ and $V$. Recall also that
\begin{equation}\label{eq:subspace.dimension-equality}
  \dim (U + V) = \dim U + \dim V - \dim (U \cap V).
\end{equation}

We will make extensive use of the fact that
\begin{equation}\label{eq:stacked-matrices-subspace-sum}
  \linspan{\mat{X \\ Y}} = \linspan{X} + \linspan{Y}
\end{equation}
and therefore
\begin{align}
\rank \mat{X \\ Y} &= \dim(\linspan{X} + \linspan{Y}) \nonumber \\
&= \rank X + \rank Y - \dim (\linspan{X} \cap \linspan{Y}). \label{eq:rank.stacked}
\end{align}

\subsection{Rank-Metric Codes}
\label{sec:rank-metric-codes}

A \defn{matrix code} is defined as any nonempty subset of $\Fq^{n\times m}$.
A matrix code is also commonly known as an \defn{array code} when it forms
a linear space over $\Fq$ \cite{Roth1991:MaximumRankArrayCodes}.

A natural and useful distance measure between elements of $\Fq^{n \times m}$ is given in the following definition.

\medskip
\begin{definition}\label{def:rank-distance}
  For $X,Y \in \Fq^{n\times m}$, the \defn{rank distance} between $X$ and $Y$
  is defined as $\dr(X,Y) \triangleq \rank(Y - X)$.
\end{definition}
\medskip

As observed in \cite{Gabidulin1985}, rank distance is
indeed a \emph{metric}.  In particular,
the triangle inequality for the rank metric follows
directly from (\ref{eq:rank.triangle-inequality}).
In the context of the rank metric, a matrix code is called a
\defn{rank-metric code}. The minimum (rank) distance of a
rank-metric code $\mathcal{C} \subseteq \Fq^{n\times m}$ is defined as
\begin{equation}\nonumber
  \dr(\mathcal{C}) \triangleq \min_{\substack{\bm{x},\bm{x}' \in \mathcal{C} \\ \bm{x} \neq \bm{x}'}} \dr(\bm{x},\bm{x}').
\end{equation}

Associated with every rank-metric code
$\mathcal{C} \subseteq \Fq^{n\times m}$
is the \defn{transposed code} $\mathcal{C}^T \subseteq \Fq^{m \times n}$,
whose codewords are obtained by transposing the codewords
of $\mathcal{C}$, i.e.,
$\mathcal{C}^T = \{ \bm{x}^T: \bm{x} \in \mathcal{C} \}$.
We have $|\mathcal{C}^T| = |\mathcal{C}|$ and
$\dr(\mathcal{C}^T) = \dr(\mathcal{C})$.
Observe the symmetry between rows and columns in the rank metric;
the distinction between a code and its transpose
is in fact transparent to the metric.

A minimum distance decoder for a rank-metric code $\mathcal{C} \subseteq \Fq^{n \times m}$
takes a word $\bm{r} \in \Fq^{n \times m}$ and returns a codeword
$\hat{\bm{x}} \in \mathcal{C}$ that is closest to $\bm{r}$ in rank distance, that is,
\begin{equation}\label{eq:rank-decoding.standard}
  \hat{\bm{x}} = \argmin_{\bm{x} \in \mathcal{C}}\, \rank(\bm{r} - \bm{x}).
\end{equation}
Note that if $\dr(\bm{x},\bm{r}) < \dr(\mathcal{C})/2$, then a minimum distance decoder is guaranteed to return $\hat{\bm{x}} = \bm{x}$.

Throughout this paper, problem (\ref{eq:rank-decoding.standard})
will be referred to as the \emph{conventional} rank decoding problem.

There is a rich coding theory for rank-metric codes that is analogous
to the classical coding theory in the Hamming metric. In particular,
we mention the existence of a Singleton bound \cite{Gabidulin1985,Delsarte1978:BilinearForms} (see also \cite{Loidreau2001:Thesis}\cite{Gadouleau.Yan2006}),
which states that every rank metric code
$\mathcal{C} \subseteq \Fq^{n \times m}$ with minimum distance
$d = \dr(\mathcal{C})$ must satisfy
\begin{align}
\log_q |\mathcal{C}|
&\leq \min \{n(m-d+1),\, m(n-d+1) \} \nonumber \\
&= \max\{n,m\} (\min\{n,m\} - d + 1). \label{eq:rankmetric.singleton-bound}
\end{align}
Codes that achieve this bound are called \defn{maximum-rank-distance}
(MRD) codes. An extensive class of MRD codes with $n\leq m$ was presented by Gabidulin in
\cite{Gabidulin1985}. By transposition, MRD codes with $n > m$ can also be obtained. Thus, MRD codes exist for all $n$ and $m$ and all $d \leq \min\{n,m\}$, irrespectively of the field size $q$.

\subsection{Subspace Codes}
\label{sec:subspace-codes}

Let $\PP(\Fq^M)$ denote the set of all subspaces of $\Fq^M$. We review
some concepts of the coding theory for subspaces developed
in \cite{Kotter.Kschischang2008}.

\medskip
\begin{definition}\label{def:subspace-distance}
Let $V,V' \in \PP(\Fq^M)$. The \defn{subspace distance} between
$V$ and $V'$ is defined as
\begin{align}
  \ds(V,V') &\triangleq \dim(V + V') - \dim(V \cap V')  \nonumber \\
   &= 2\dim(V + V') - \dim V -\dim V'  \label{eq:subspace.dist.a} \\
   &= \dim V + \dim V' - 2\dim(V \cap V'). \label{eq:subspace.dist.b}
  \end{align}
\end{definition}
\medskip

It is shown in \cite{Kotter.Kschischang2008} that the subspace distance
is indeed a metric on $\PP(\Fq^M)$.

A \defn{subspace code} is defined as a nonempty subset of $\PP(\Fq^M)$.
The minimum (subspace) distance of a subspace code
$\Omega \subseteq \PP(\Fq^M)$ is defined as
\begin{equation}\nonumber
  \ds(\Omega) \triangleq \min_{\substack{V,V' \in \Omega \\ V \neq V'}} \ds(V,V').
\end{equation}

The minimum distance decoding problem for a subspace code is to find a
subspace $\hat{V} \in \Omega$ that is closest to a given subspace
$U \in \PP(\Fq^M)$, i.e.,
\begin{equation}\label{eq:subspace-decoding}
  \hat{V} = \argmin_{V \in \Omega}\, \ds(V,U).
\end{equation}
A minimum distance decoder is guaranteed to return $\hat{V} = V$ if $\ds(V,U) < \ds(\Omega)/2$.

Let $\PP(\Fq^M, n)$ denote the set of all $n$-dimensional subspaces
of $\Fq^M$. A subspace code $\Omega$ is called a constant-dimension
code if $\Omega \subseteq \PP(\Fq^M, n)$. It follows
from (\ref{eq:subspace.dist.a}) or (\ref{eq:subspace.dist.b})
that the minimum distance
of a constant-dimension code is always an even number.

Let $A_q[M,2d,n]$ be denote the maximum number of codewords in a constant-dimension code with minimum subspace distance $2d$. Many bounds on $A_q[M,2d,n]$ were developed in \cite{Kotter.Kschischang2008}, in particular the Singleton-like bound
\begin{equation}\label{eq:subspace.singleton-bound}
  A_q[M,2d,n] \leq \mat{M - d + 1 \\ \max\{n,M-n\}}_q
\end{equation}
where
\[
{{M}\brack{n}}_q  \triangleq
\frac{(q^{M}-1) \cdots (q^{M-n+1}-1)}{(q^{n}-1) \cdots (q-1)}
\]
denotes the \defn{Gaussian coefficient}. It is well known that the
Gaussian coefficient gives the number of distinct $n$-dimensional subspaces
of an $M$-dimensional vector space over $\Fq$, i.e., ${{M}\brack{n}}_q = |\PP(\Fq^M, n)|$. A useful bound on ${{M}\brack{n}}_q$ is given by
\cite[Lemma 5]{Kotter.Kschischang2008}
\begin{equation}\label{eq:Gaussian-coefficient.bound}
  {{M}\brack{n}}_q < 4q^{n(M-n)}.
\end{equation}
Combining (\ref{eq:subspace.singleton-bound}) and (\ref{eq:Gaussian-coefficient.bound}) gives
\begin{equation}\label{eq:subspace.singleton-bound.approx}
  A_q[M,2d,n] < 4 q^{\max\{n,M-n\} (\min\{n,M-n\} - d + 1)}.
\end{equation}

There exist also bounds on $A_q[M,2d,n]$ that are tighter than (\ref{eq:subspace.singleton-bound}), namely the Wang-Xing-Safavi-Naini bound \cite{Wang++2003} and a Johnson-type bound \cite{Xia2007}.

For future reference, we define the \defn{sub-optimality} of a constant-dimension code $\Omega \subseteq \PP(\Fq^M, n)$ with $\ds(\Omega) = 2d$ to be
\begin{equation}\label{eq:sub-optimality}
  \alpha(\Omega) \triangleq \frac{\log_q A_q[M,2d,n] - \log_q |\Omega|}{\log_q A_q[M,2d,n]}.
\end{equation}

\section{Error Control in Random Network Coding}
\label{sec:random-network-coding}

\subsection{Channel Model}
\label{sec:channel-model}

We start by reviewing the basic model for single-source
generation-based random linear network coding \cite{Ho++2006,Chou++2003}.
Consider a point-to-point communication network with a single source node
and a single destination node.
Each link in the network is assumed to
transport, free of errors, a packet of $M$ symbols in a finite field $\Fq$.
Links are directed, incident \emph{from} the node transmitting the packet
and incident \emph{to} the node receiving the packet.
A packet transmitted on a link incident to a given node is said to be
an \defn{incoming packet} for that node, and similarly a packet transmitted
on a link incident from a given node is said to be an \defn{outgoing packet}
for that node.

During each transmission generation,
the source node formats the information to be transmitted into $n$ packets
$X_1,\ldots,X_n \in \Fq^{1 \times M}$, which are regarded as incoming packets
for the source node. Whenever a node (including the source) has a transmission
opportunity, it produces an outgoing packet as a random $\Fq$-linear
combination of all the incoming packets it has until then received.
The destination node collects $N$ packets
$Y_1,\ldots,Y_{N} \in \Fq^{1 \times M}$ and tries to recover the original
packets $X_1,\ldots,X_n$.

Let $X$ be an $n \times M$ matrix whose rows are the transmitted packets
$X_1,\ldots,X_n$ and, similarly, let $Y$ be an $N \times M$ matrix whose
rows are the received packets $Y_1,\ldots,Y_{N}$. Since all packet operations
are linear over $\Fq$, then, regardless of the network topology, the
transmitted packets $X$ and the received packets $Y$ can be related as
\begin{equation}\label{eq:basic-channel}
 Y = AX,
\end{equation}
where $A$ is an $N \times n$  matrix corresponding to the overall
linear transformation applied by the network.

Before proceeding, we remark that this model encompasses a variety
of situations:
\begin{itemize}
\item The network may have cycles or delays. Since the overall system
is linear, expression (\ref{eq:basic-channel}) will be true regardless of
the network topology.
\item The network could be wireless instead of wired. Broadcast transmissions
in wireless networks
may be modeled by
constraining each intermediate node to send exactly the same packet
on each of its outgoing links.
\item The source node may transmit more than one \defn{generation}
(a set of $n$ packets). In this case, we assume that each packet carries
a label identifying the generation to which it corresponds and that
packets from different generations are processed separately throughout
the network \cite{Chou++2003}.
\item The network topology may be time-varying as nodes join and leave and
connections are established and lost. In this case, we assume that each
network link is the instantiation of an actual successful packet transmission.
  \item The network may be used for multicast, i.e., there may be more than
one destination node. Again, expression (\ref{eq:basic-channel}) applies;
however, the matrix $A$ may be different for each destination.
\end{itemize}

Let us now extend this model to incorporate packet errors.
Following \cite{Cai.Yeung2002,Yeung.Cai2006,Cai.Yeung2006}, we consider that
packet errors may occur in any of the links of the network.
Suppose the links in the network are indexed from 1 to $\nlinks$,
and let $Z_i$ denote the error packet applied at link
$i \in \{1,\ldots,\nlinks\}$.  The application of an error packet is modeled
as follows.
We assume that, for each link $i$, the node transmitting on that link
first creates a prescribed packet $P_{\text{in},i} \in \Fq^{1 \times M}$
following the procedure described above. Then, an error packet
$Z_i \in \Fq^{1 \times M}$ is added to $P_{\text{in},i}$ in order
to produce the outgoing packet on this link, i.e.,
$P_{\text{out},i} = P_{\text{in},i} + Z_i$.
Note that any arbitrary packet $P_{\text{out},i}$ can
be formed simply by choosing $Z_i = P_{\text{out},i} - P_{\text{in},i}$.

Let $Z$ be an $\nlinks \times M$ matrix whose rows are the error
packets $Z_1,\ldots,Z_\nlinks$. By linearity of the network,
we can write
\begin{equation}\label{eq:channel-model}
  Y = AX + BZ,
\end{equation}
where $B$ is an $N \times \nlinks$ matrix corresponding to the overall
linear transformation applied to $Z_1,\ldots,Z_\nlinks$ on route to the
destination. Note that $Z_i =0$ means that no corrupt packet was injected
at link $i$. Thus, the number of nonzero rows of $Z$, $\wt(Z)$, gives the
total number of (potentially) corrupt packets injected in the network.
Note that it is possible that a nonzero error packet happens to be in the row space of $X$, in which case it is not really a corrupt packet.

Observe that this model can represent not only the occurrence of random
link errors, but also the action of malicious nodes. A malicious node can
potentially transmit erroneous packets on all of its outgoing links.
A malicious node may also want to disguise itself and transmit correct
packets on some of these links, or may simply refuse to transmit some
packet (i.e., transmitting an all-zero packet), which is represented in
the model by setting $Z_i = - P_{\text{in},i}$. In any case, $\wt(Z)$ gives
the total number of ``packet interventions'' performed by all malicious nodes
and thus gives a sense of the total adversarial ``power'' employed towards
jamming the network.

Equation (\ref{eq:channel-model}) is our basic model of a channel induced by
random linear network coding, and we will refer to it as the
\defn{random linear network coding channel} (RLNCC). The channel input and output
alphabets are given by $\Fq^{n \times M}$ and $\Fq^{N \times M}$, respectively.
To give a full probabilistic specification of the channel, we would need
to specify the joint probability distribution of $A$, $B$ and $Z$ given $X$.
We will not pursue this path in this paper, taking, instead,
a more combinatorial approach.

\subsection{Transmission via Subspace Selection}
\label{sec:transmission-via-subspace-selection}

Let $\Omega \subseteq \PP(\Fq^M)$ be a subspace code with maximum dimension $n$. In the approach in \cite{Kotter.Kschischang2008}, the source node selects a subspace $V \in \Omega$ and transmits this subspace over the RLNCC as some matrix $X \in \Fq^{n \times M}$ such that $V = \linspan{X}$. The destination node receives $Y \in \Fq^{N \times M}$ and computes $U = \linspan{Y}$, from which the transmitted subspace can be inferred using a minimum distance decoder (\ref{eq:subspace-decoding}).

In this paper, it will be convenient to view the above approach from a matrix perspective. In order to do that, we simply replace $\Omega$ by an (arbitrarily chosen) matrix code that generates $\Omega$. More precisely, let $[\Omega] \triangleq \{X \in \Fq^{n \times M} \colon X = \RRE(X),\, \linspan{X} \in \Omega\}$ be a matrix code consisting of all the $n \times M$ matrices in RRE form whose row space is in $\Omega$. Now, the above setup can be reinterpreted as follows. The source node selects a matrix $X \in [\Omega]$ to transmit over the RLNCC. Upon reception of $Y$, the destination node tries to infer the transmitted matrix using the minimum distance decoding rule
\begin{equation}\label{eq:RLNCC.decoding-rule}
  \hat{X} = \argmin_{X \in [\Omega]}\, \ds(\linspan{X},\linspan{Y}).
\end{equation}
Note that the decoding is guaranteed to be successful if $\ds(\linspan{X},\linspan{Y}) < \ds(\Omega)/2$.

\subsection{Performance Guarantees}
\label{sec:guarantees}

In this subsection, we wish to relate the performance guarantees of a subspace code with more concrete network parameters. Still, we would like these parameters to be sufficiently general so that we do not need to take the whole network topology into account.

We make the following assumptions:

\begin{itemize}
  \item The column-rank deficiency of the transfer matrix $A$ is never greater than $\rho$, i.e., $\rank A \geq n - \rho$.
  \item The adversarial nodes together can inject at most $t$ corrupting packets, i.e., $\wt(Z) \leq t$.
\end{itemize}

The following result characterizes the performance guarantees of a subspace code under our assumptions.

\medskip
\begin{theorem}\label{thm:RLNCC-code.capability}
  Suppose $\rank A \geq n - \rho$ and $\wt(Z) \leq t$. Then, decoding according to (\ref{eq:RLNCC.decoding-rule}) is guaranteed to be successful provided
  $2t + \rho < \ds(\Omega)/2$.
\end{theorem}
\medskip

In order to prove Theorem~\ref{thm:RLNCC-code.capability}, we need a few results relating rank and subspace distance.

\medskip
\begin{proposition}\label{prop:rank-stacked.bound}
  Let $X,Y \in \Fq^{N \times M}$. Then
  \begin{equation}\nonumber
    \rank \mat{X \\ Y} \leq \rank(Y - X) + \min\{\rank X,\, \rank Y\}.
  \end{equation}
\end{proposition}
\begin{IEEEproof}
  We have
  \begin{align}
    \rank \mat{X \\ Y} &= \rank \mat{X \\ Y-X} \leq \rank(Y-X) + \rank X  \nonumber \\
    \rank \mat{X \\ Y} &= \rank \mat{Y-X \\ Y} \leq \rank(Y-X) + \rank Y. \nonumber
  \end{align}
\end{IEEEproof}
\medskip

\begin{corollary}\label{cor:subspace-distance.bound}
  Let $X,Z \in \Fq^{N \times M}$ and $Y = X + Z$. Then
  \begin{equation}\nonumber
    \ds(\linspan{X},\linspan{Y}) \leq 2 \rank Z - |\rank X - \rank Y|.
  \end{equation}
\end{corollary}
\begin{IEEEproof}
  From Proposition~\ref{prop:rank-stacked.bound}, we have
  \begin{align}
  \ds(\linspan{X},\linspan{Y})
  &= 2\rank \mat{X \\ Y} - \rank X - \rank Y \nonumber \\
  &\leq 2\rank Z + 2\min\{\rank X,\, \rank Y\} \nonumber \\
  &\quad\, - \rank X - \rank Y \nonumber \\
  &= 2\rank Z - |\rank X - \rank Y|. \nonumber
  \end{align}
\end{IEEEproof}
\medskip

We can now give a proof of Theorem~\ref{thm:RLNCC-code.capability}.

\medskip
\begin{IEEEproof}[Proof of Theorem~\ref{thm:RLNCC-code.capability}]
  From Corollary~\ref{cor:subspace-distance.bound}, we have that
  \begin{equation}\nonumber
    \ds(\linspan{AX},\linspan{Y}) \leq 2 \rank BZ \leq 2 \rank Z \leq 2 \wt(Z) \leq 2 t.
  \end{equation}
  Using (\ref{eq:rank.product-inequality}), we find that
  \begin{equation}\nonumber
    \ds(\linspan{X},\linspan{AX}) =  \rank X - \rank AX \leq n - \rank A \leq \rho.
  \end{equation}
  Since $\ds(\cdot,\cdot)$ satisfies the triangle inequality, we have
  \begin{align}
  \ds(\linspan{X},\linspan{Y})
  &\leq \ds(\linspan{X},\linspan{AX}) + \ds(\linspan{AX},\linspan{Y}) \nonumber \\
  &\leq \rho + 2t \nonumber \\
  &< \frac{\ds(\Omega)}{2} \nonumber
  \end{align}
  and therefore the decoding is guaranteed to be successful.
\end{IEEEproof}
\medskip

Theorem~\ref{thm:RLNCC-code.capability} is analogous to Theorem 2 in \cite{Kotter.Kschischang2008}, which states that minimum subspace distance decoding is guaranteed to be successful if $2(\mu + \delta) < \ds(\Omega)$, where $\delta$ and $\mu$ are, respectively, the number of ``insertions'' and ``deletions'' of dimensions that occur in the channel \cite{Kotter.Kschischang2008}. Intuitively, since one corrupted packet injected at a network min-cut can effectively replace a dimension of the transmitted subspace, we see that $t$ corrupted packets can cause $t$ deletions and $t$ insertions of dimensions. Combined with possible $\rho$ further deletions caused by a row-rank deficiency of $A$, we have that $\delta = t$ and $\mu = t + \rho$. Thus,
  \begin{equation}\nonumber
    \delta + \mu < \frac{\ds(\Omega)}{2} \implies 2t + \rho < \frac{\ds(\Omega)}{2}.
  \end{equation}
In other words, under the condition that corrupt packets may be injected in any of the links in network (which must be assumed if we do not wish to take the network topology into account), the performance guarantees of a minimum distance decoder are essentially given by Theorem~\ref{thm:RLNCC-code.capability}.

It is worth to mention that, according to recent results \cite{Silva.Kschischang2008:Metrics},
minimum subspace distance decoding may not be the optimal decoding rule
when the subspaces in $\Omega$ have different dimensions. For the remainder
of this paper, however, we focus on the case of a constant-dimension code
and therefore we use the minimum distance decoding
rule (\ref{eq:RLNCC.decoding-rule}). Our goal will be to construct constant-dimension subspace
codes with good performance and efficient encoding/decoding procedures.

\section{Codes for the Random Linear Network Coding Channel Based on Rank-Metric Codes}
\label{sec:rank-metric-approach}

In this section, we show how a constant-dimension subspace code
can be constructed from any rank-metric code. In particular, this
construction will allow us to obtain nearly-optimal subspace codes that
possess efficient encoding and decoding algorithms.

\subsection{Lifting Construction}
\label{sec:lifting}

From now on, assume that $M = n + m$, where $m>0$. Let $I = I_{n \times n}$.

\medskip
\begin{definition}\label{def:lifting-construction}
Let $\I\colon \Fq^{n \times m} \to \PP(\Fq^{n + m})$, given by $\bm{x} \mapsto \I(\bm{x}) = \linspan{\mat{I & \bm{x}}}$. The subspace $\I(\bm{x})$ is called the \defn{lifting} of the matrix $\bm{x}$. Similarly, if $\mathcal{C} \subseteq \Fq^{n \times m}$ is a rank-metric code, then the subspace code $\I(\mathcal{C})$, obtained by lifting each codeword of $\mathcal{C}$, is called the \defn{lifting} of $\mathcal{C}$.
\end{definition}
\medskip

Definition~\ref{def:lifting-construction} provides an injective mapping between rank-metric codes and subspace codes. Note that a subspace code constructed by lifting is always a constant-dimension code (with codeword dimension $n$).

Although the lifting construction is a particular way of
constructing subspace codes, it can also be seen as a generalization
of the standard approach to random network coding \cite{Ho++2006,Chou++2003}.
In the latter, every transmitted matrix has the form
$X = [I \quad \bm{x}]$, where the payload matrix $\bm{x} \in \Fq^{n \times m}$
corresponds to the raw data to be communicated. In our approach,
each transmitted matrix is also of the form $X = [I \quad \bm{x}]$,
but the payload matrix $\bm{x} \in \mathcal{C}$ is restricted to be a codeword
of a rank-metric code rather than uncoded data.

Our reasons for choosing $\mathcal{C}$ to be a rank-metric code will
be made clear from the following proposition.

\medskip
\begin{proposition}\label{prop:rank-distance-and-subspace-distance}
Let $\mathcal{C} \subseteq \Fq^{n \times m}$ and $\bm{x},\bm{x}' \in \mathcal{C}$. Then
\begin{align}
  \ds(\I(\bm{x}),\I(\bm{x}')) &= 2\dr(\bm{x},\bm{x}') \nonumber \\
  \ds(\I(\mathcal{C})) &= 2\dr(\mathcal{C}). \nonumber
  \end{align}
\end{proposition}
\begin{IEEEproof} Since $\dim \I(\bm{x}) = \dim \I(\bm{x}') = n$, we have
\begin{align}
  \ds(\I(\bm{x}),\I(\bm{x}'))
  &= 2\dim(\I(\bm{x}) + \I(\bm{x}')) -2n %-\dim \I(\bm{x})-\dim \I(\bm{x}')
  \nonumber \\
  &= 2\rank \mat{I & \bm{x} \\ I & \bm{x}'} - 2n \nonumber \\
  &= 2\rank \mat{I & \bm{x} \\ 0 & \bm{x}' - \bm{x}} - 2n \nonumber \\
  &= 2\rank (\bm{x}'-\bm{x}). \nonumber
\end{align}
The second statement is immediate.
\end{IEEEproof}
\medskip

Proposition~\ref{prop:rank-distance-and-subspace-distance} shows that a
subspace code constructed by lifting inherits the distance properties
of its underlying rank-metric code. The question of whether such lifted
rank-metric codes are ``good'' compared to the whole class of
constant-dimension codes is addressed in the following proposition.

\medskip
\begin{proposition}\label{prop:MRD-near-optimal}
  Let $\mathcal{C} \subseteq \Fq^{n \times m}$ be an MRD code with $\dr(\mathcal{C}) = d$. Then $\ds(\I(\mathcal{C})) = 2d$ and
\begin{equation}\nonumber
  A_q[n+m,2d,n] < 4 |\I(\mathcal{C})| = 4 |\mathcal{C}|.
\end{equation}
Moreover, for any code parameters, the sub-optimality of $\I(\mathcal{C})$ in $\PP(\Fq^{n+m},n)$ satisfies
\begin{equation}\nonumber
  \alpha(\I(\mathcal{C})) < \frac{4}{(n+m)\log_2 q}.
\end{equation}
\end{proposition}
\begin{IEEEproof}
Using (\ref{eq:subspace.singleton-bound.approx}) and the fact that $\mathcal{C}$ achieves the Singleton bound for rank-metric codes (\ref{eq:rankmetric.singleton-bound}), we have
\begin{align}
  A_q[n+m,2d,n]
  &< 4 q^{\max\{n,m\} (\min\{n,m\} - d + 1)} \nonumber \\
  &= 4 |\mathcal{C}|. \nonumber
\end{align}
Applying this result in (\ref{eq:sub-optimality}),
we obtain
\begin{align}
  \alpha(\I(\mathcal{C}))
  &< \frac{\log_q 4}{\max\{n,m\}(\min\{n,m\}-d+1)} \nonumber \\
  &\leq \frac{\log_q 4}{\max\{n,m\}} \nonumber \\
  &\leq \frac{\log_q 4}{(n+m)/2} \nonumber \\
  &= \frac{4}{(n+m)\log_2 q}. \nonumber
\end{align}
\end{IEEEproof}
\medskip

Proposition~\ref{prop:MRD-near-optimal} shows that, for all practical purposes, lifted MRD codes are essentially optimal as constant-dimension codes. Indeed, the rate loss in using a lifted MRD code rather than an optimal constant-dimension code is smaller than $4/P$, where $P = (n+m)\log_2 q$ is the packet size in bits. In particular, for packet sizes of 50 bytes or more, the rate loss is smaller than 1\%.

In this context, it is worth mentioning that the
nearly-optimal Reed-Solomon-like codes proposed in
\cite{Kotter.Kschischang2008} correspond exactly to the lifting
of the class of MRD codes proposed by
Gabidulin \cite{Gabidulin1985}. The latter will be discussed in more detail in Section~\ref{sec:decoding-error-erasures-deviations}.

\subsection{Decoding}
\label{sec:decoding-lifted-codes}

We now specialize the decoding problem (\ref{eq:RLNCC.decoding-rule}) to the specific case of lifted rank-metric codes.
We will see that it is possible to reformulate such a problem
in a way that resembles the conventional decoding problem for
rank-metric codes, but with additional side-information presented
to the decoder.

Let the transmitted matrix be given by $X = [I \quad \bm{x}]$,
where $\bm{x} \in \mathcal{C}$ and $\mathcal{C} \subseteq \Fq^{n \times m}$
is a rank-metric code.  Write the received matrix as
\begin{equation}\nonumber
  Y = [\hat{A} \quad \bm{y}]
\end{equation}
where $\hat{A} \in \Fq^{N \times n}$ and $\bm{y} \in \Fq^{N \times m}$.
In accordance with the formulation of Section~\ref{sec:transmission-via-subspace-selection},
we assume that $\rank Y = N$, since any linearly dependent received packets
do not affect the decoding problem and may be discarded by the destination
node. Now, define
\begin{equation}\nonumber
  \mu \triangleq n - \rank \hat{A} \quad \text{ and } \quad \delta \triangleq N - \rank \hat{A}.
\end{equation}
Here $\mu$ measures the rank deficiency of $\hat{A}$ with respect to columns,
while $\delta$ measures the rank deficiency of $\hat{A}$ with respect to rows.

Before examining the general problem, we study the simple
special case that arises when $\mu = \delta = 0$.

\medskip
\begin{proposition}\label{prop:decoding-lifted-codes.special-case}
  If $\mu = \delta = 0$, then
  \begin{equation}\nonumber
    \ds(\linspan{X},\linspan{Y}) = 2\dr(\bm{x},\bm{r})
  \end{equation}
  where $\bm{r} = \hat{A}^{-1} \bm{y}$.
\end{proposition}
\begin{IEEEproof}
  Since $\mu = \delta = 0$, $\hat{A}$ is invertible. Thus, $\bar{Y} = [I \quad \hat{A}^{-1} \bm{y}]$ is row equivalent to $Y$, i.e., $\linspan{\bar{Y}} = \linspan{Y}$. Applying Proposition~\ref{prop:rank-distance-and-subspace-distance}, we get the desired result.
\end{IEEEproof}
\medskip

The above proposition shows that, whenever $\hat{A}$ is invertible,
a solution to (\ref{eq:RLNCC.decoding-rule}) can be found by solving
the conventional rank decoding problem.  This case is illustrated
by the following example.

\begin{example}\label{ex:decoding-with-errors}
  Let $n=4$ and $q=5$. Let $x_1,\ldots,x_4$ denote the rows of a codeword $\bm{x} \in \mathcal{C}$. Suppose that
  \begin{equation}\nonumber
    A = \mat{2 & 4 & 2 & 4 \\ 0 & 0 & 3 & 3 \\ 1 & 0 & 4 & 3 \\ 0 & 4 & 1 & 4},
  \end{equation}
  $B = \mat{4 & 0 & 1 & 0}^T$ and $Z = \mat{1 & 2 & 3 & 4 & z}$. Then
  \begin{equation}\nonumber
    Y = \mat{1 & 2 & 4 & 0 & 2x_1+4x_2+2x_3+4x_4+4z \\ 0 & 0 & 3 & 3 & 3x_3+3x_4 \\ 2 & 2 & 2 & 2 & x_1+4x_3+3x_4+z \\ 0 & 4 & 1 & 4 & 4x_2+x_3+4x_4}.
  \end{equation}
  Converting $Y$ to RRE form, we obtain
  \begin{equation}\label{eq:reduced-Y-example-1}
    \bar{Y} = \mat{I & \bm{r}}
  \end{equation}
  where
  \begin{align}
  \bm{r} &= \mat{3x_2+2x_3+x_4+z \\ 3x_1+2x_2+4x_3+2x_4+2z \\ 4x_1+3x_2+3x_3+x_4+z \\ x_1+2x_2+3x_3+4z}. \nonumber
  \end{align}
  Note that, if no errors had occurred, we would expect to find $\bm{r} = \bm{x}$.

  Now, observe that we can write
  \begin{align}
  \bm{r} &= \mat{x_1 \\ x_2 \\ x_3 \\ x_4} + \mat{1 \\ 2 \\ 1 \\ 4}\mat{4x_1+3x_2+2x_3+x_4+z}. \nonumber
  \end{align}
Thus, $\rank (\bm{r} - \bm{x}) = 1$. We can think of this
as an error word $\bm{e} = \bm{r} - \bm{x}$ of rank~1 applied to $\bm{x}$.
This error can be corrected if $\dr(\mathcal{C}) \geq 3$.
\qed
\end{example}
\medskip

Let us now proceed to the general case, where $\hat{A}$ is not necessarily
invertible. We first examine a relatively straightforward approach that,
however, leads to an unattractive decoding problem.

Similarly to the
proof of Proposition~\ref{prop:decoding-lifted-codes.special-case},
it is possible to show that
\begin{equation}\nonumber
  \ds(\linspan{X},\linspan{Y}) = 2\rank(\bm{y} - \hat{A}\bm{x}) + \mu - \delta
\end{equation}
which yields the following decoding problem:
\begin{equation}\label{eq:rankmetric.decoding.bad}
  \hat{\bm{x}} = \argmin_{\bm{x} \in \mathcal{C}} \, \rank(\bm{y}-\hat{A}\bm{x}).
\end{equation}
If we define a new code
$\mathcal{C}' = \hat{A}\mathcal{C}=\{\hat{A}\bm{x},\,\bm{x}\in\mathcal{C}\}$,
then a solution to (\ref{eq:rankmetric.decoding.bad}) can be found by
first solving
\begin{equation}\nonumber
  \hat{\bm{x}}' = \argmin_{\bm{x}' \in \mathcal{C}'} \, \rank (\bm{y} - \bm{x}')
\end{equation}
using a conventional rank decoder for $\mathcal{C}'$ and then choosing
any $\hat{\bm{x}} \in \{\bm{x} \mid \hat{A}\bm{x} = \hat{\bm{x}}'\}$ as
a solution.
An obvious drawback of this approach is that it requires a new code
$\mathcal{C}'$ to be used at each decoding instance.
This is likely to increase the decoding complexity, since the existence
of an efficient algorithm for $\mathcal{C}$ does not imply the existence
of an efficient algorithm for $\mathcal{C}'= \hat{A} \mathcal{C}$ for all
$\hat{A}$. Moreover, even if efficient algorithms are known for all
$\mathcal{C}'$, running a different algorithm for each received matrix
may be impractical or undesirable from an implementation point-of-view.

In the following, we seek an expression for $\ds(\linspan{X},\linspan{Y})$ where the
structure of $\mathcal{C}$ can be exploited. In order to motivate
our approach, we consider the following two examples, which
generalize Example~\ref{ex:decoding-with-errors}.

\medskip
\begin{example}
Let us return to Example~\ref{ex:decoding-with-errors}, but now suppose
  \begin{equation}\nonumber
    A = \mat{1 & 0 & 2 & 3 \\ 1 & 3 & 0 & 3 \\ 1 & 4 & 0 & 3 \\ 2 & 0 & 4 & 0 \\ 1 & 1 & 2 & 4},
  \end{equation}
  $B = \mat{4 & 0 & 1 & 0 & 0}^T$ and $Z = \mat{1 & 2 & 3 & 4 & z}$. Then
  \begin{equation}\nonumber
    Y = \mat{0 & 3 & 4 & 4 & x_1+2x_3+3x_4+4z \\ 1 & 3 & 0 & 3 & x_1+3x_2+3x_4 \\ 2 & 1 & 3 & 2 & x_1+4x_2+3x_4+z \\ 2 & 0 & 4 & 0 & 2x_1+4x_3 \\ 1 & 1 & 2 & 4 & x_1+x_2+2x_3+4x_4} = \mat{\hat{A} & \bm{y}}.
  \end{equation}
Although $\hat{A}$ is not invertible, we can nevertheless convert $Y$ to RRE form to obtain
  \begin{equation}\label{eq:reduced-Y-example-2}
    \bar{Y} = \mat{I & \bm{r} \\ 0 & \hat{E}}
  \end{equation}
  where
  \begin{equation}\nonumber
     \bm{r} = \mat{2x_1+2x_2+3x_3+4x_4+4z \\ 4x_1+4x_2+2x_3+x_4+z \\ 2x_1+4x_2+2x_3+3x_4+3z \\ 3x_1+x_2+4x_3+3x_4+2z}
  \end{equation}
  and
  \begin{equation}\nonumber
    \hat{E} = 2x_1+4x_2+x_3+3x_4+3z.
  \end{equation}
  Observe that
  \begin{equation}\nonumber
    \bm{e} = \bm{r} - \bm{x} = \mat{x_1+2x_2+3x_3+4x_4+4z \\ 4x_1+3x_2+2x_3+x_4+z \\ 2x_1+4x_2+x_3+3x_4+3z \\ 3x_1+x_2+4x_3+2x_4+2z} = \mat{3 \\ 2 \\ 1 \\ 4}\hat{E}.
  \end{equation}
Thus, we see not only that $\rank \bm{e} = 1$, but we have also recovered
part of its decomposition as an outer product, namely, the vector $\hat{E}$.
\qed
\end{example}
\medskip

\begin{example}
Consider again the parameters of Example~\ref{ex:decoding-with-errors},
but now let
  \begin{equation}\nonumber
    A = \mat{3 & 2 & 1 & 1 \\ 0 & 4 & 3 & 2 \\ 2 & 1 & 0 & 4}
  \end{equation}
and suppose that there are no errors. Then
  \begin{equation}\nonumber
    Y = \mat{3 & 2 & 1 & 1 & 3x_1+2x_2+x_3+x_4 \\ 0 & 4 & 3 & 2 & 4x_2+3x_3+2x_4 \\ 2 & 1 & 0 & 4 & 2x_1+x_2+4x_4} = \mat{\hat{A} & \bm{y}}.
  \end{equation}
Once again we cannot invert $\hat{A}$; however, after converting $Y$ to RRE form and inserting an all-zero row
in the third position, we obtain
\begin{align}
  \hat{Y} &= \mat{1 & 0 & 4 & 0 & x_1+4x_3 \\ 0 & 1 & 2 & 0 & x_2+2x_3 \\ 0 & 0 & 0 & 0 & 0 \\ 0 & 0 & 0 & 1 & x_4} \nonumber \\
  &= \mat{1 & 0 & 4 & 0 & x_1+4x_3 \\ 0 & 1 & 2 & 0 & x_2+2x_3 \\ 0 & 0 & 1 - 1 & 0 & x_3 - x_3 \\ 0 & 0 & 0 & 1 & x_4} \nonumber \\
  &= \mat{I + \hat{L} I_3^T & \bm{x} + \hat{L}x_3} \nonumber \\
  &= \mat{I + \hat{L} I_3^T & \bm{r}} \label{eq:reduced-Y-example-3}
  \end{align}
where
  \begin{equation}\nonumber
    \hat{L} = \mat{4 \\ 2 \\ -1 \\ 0}.
  \end{equation}
Once again we see that
the error word has rank 1, and that we have recovered
part of its decomposition as an outer product. Namely, we have
  \begin{equation}\nonumber
    \bm{e} = \bm{r} - \bm{x} = \hat{L}x_3
  \end{equation}
where this time $\hat{L}$ is known.
\qed
\end{example}
\medskip

Having seen from these two examples
how side information (partial knowledge of the error matrix)
arises at the output of the RLNCC,
we address the general case in the following proposition.

\medskip
\begin{proposition}\label{prop:reduction.existence}
  Let $Y$, $\mu$ and $\delta$ be defined as above. There exist a tuple $(\bm{r}, \hat{L}, \hat{E}) \in \Fq^{n \times m} \times \Fq^{n \times \mu} \times \Fq^{\delta \times m}$ and a set $\mathcal{U} \subseteq \{1,\ldots,n\}$ satisfying
  \begin{align}
    |\mathcal{U}| &= \mu \label{eq:reduc.property-1} \\
    I_\mathcal{U}^T \bm{r} &= 0 \label{eq:reduc.property-2} \\
    I_\mathcal{U}^T \hat{L} &= - I_{\mu \times \mu} \label{eq:reduc.property-3} \\
    \rank \hat{E} &= \delta \label{eq:reduc.property-4}
  \end{align}
  such that
  \begin{equation}\label{eq:reduc.property-5}
        \linspan{\mat{I + \hat{L} I_\mathcal{U}^T & \bm{r} \\ 0 & \hat{E}}} = \linspan{Y}.
  \end{equation}
\end{proposition}
\begin{IEEEproof}
  See the Appendix.
\end{IEEEproof}
\medskip

Proposition~\ref{prop:reduction.existence} shows that every matrix
$Y$ is row equivalent to a matrix
\begin{equation}\nonumber
  \bar{Y} = \mat{I + \hat{L} I_\mathcal{U}^T & \bm{r} \\ 0 & \hat{E}}
\end{equation}
which is essentially the matrix $Y$ in reduced row echelon form. Equations (\ref{eq:reduced-Y-example-1}), (\ref{eq:reduced-Y-example-2}) and (\ref{eq:reduced-Y-example-3}) are examples of matrices in this form. We can think of the matrices
$\bm{r}$, $\hat{L}$ and $\hat{E}$ and the set $\mathcal{U}$ as providing
a compact description of the received subspace $\linspan{Y}$. The set $\mathcal{U}$
is in fact redundant and can be omitted from the description, as we show
in the next proposition.

\medskip
\begin{proposition}\label{prop:reduction.redundant-set}
  Let $(\bm{r},\hat{L},\hat{E}) \in \Fq^{n \times m} \times \Fq^{n \times \mu} \times \Fq^{\delta \times m}$ be a tuple and $\mathcal{U} \subseteq \{1,\ldots,n\}$ be a set that satisfy (\ref{eq:reduc.property-1})--(\ref{eq:reduc.property-4}). For any $\mathcal{S} \subseteq \{1,\ldots,n\}$, $T \in \Fq^{\mu \times \mu}$ and $R \in \Fq^{\delta \times \delta}$ such that $(\bm{r}, \hat{L}T, R\hat{E})$ and $\mathcal{S}$ satisfy (\ref{eq:reduc.property-1})--(\ref{eq:reduc.property-4}), we have
\begin{equation}\nonumber
        \linspan{\mat{I + \hat{L}T I_\mathcal{S}^T & \bm{r} \\ 0 & R\hat{E}}} = \linspan{\mat{I + \hat{L} I_\mathcal{U}^T & \bm{r} \\ 0 & \hat{E}}}.
  \end{equation}
\end{proposition}
\begin{IEEEproof}
  See the Appendix.
\end{IEEEproof}
\medskip

Proposition~\ref{prop:reduction.redundant-set} shows that, given a tuple $(\bm{r},\hat{L},\hat{E})$ obtained from Proposition~\ref{prop:reduction.existence}, the set $\mathcal{U}$ can be found as \emph{any} set satisfying (\ref{eq:reduc.property-1})--(\ref{eq:reduc.property-3}). Moreover, the matrix $\hat{L}$ can be multiplied on the right by any nonsingular matrix (provided that the resulting matrix satisfies (\ref{eq:reduc.property-1})--(\ref{eq:reduc.property-3}) for some $\mathcal{U}$), and the matrix $\hat{E}$ can be multiplied on the left by any nonsingular matrix; none of these operations change the subspace described by $(\bm{r},\hat{L},\hat{E})$. The notion of a concise description of a subspace $\linspan{Y}$ is captured in the following definition.

\medskip
\begin{definition}
A tuple
$(\bm{r}, \hat{L}, \hat{E}) \in \Fq^{n \times m} \times \Fq^{n \times \mu} \times \Fq^{\delta \times m}$ that satisfies (\ref{eq:reduc.property-1})--(\ref{eq:reduc.property-5}) for some $\mathcal{U} \subseteq \{1,\ldots,n\}$ is said
to be a \defn{reduction} of the matrix $Y$.
\end{definition}
\medskip

\begin{remark}
It would be enough to specify, besides the matrix $\bm{r}$, only the column space of $\hat{L}$ and the row space of $\hat{E}$ in the definition of a reduction. For simplicity we will, however, not use this notation here.
\end{remark}
\medskip

Note that if $Y$ is a lifting of $\bm{r}$, then $(\bm{r},[],[])$ is
a reduction of $Y$ (where $[]$ denotes an empty matrix).
Thus, reduction can be interpreted as the inverse of lifting.

We can now prove the main theorem of this section.

\medskip
\begin{theorem}\label{thm:reduction-and-Delta}
  Let $(\bm{r}, \hat{L}, \hat{E})$ be a reduction of $Y$. Then
  \begin{align}
    \ds(\linspan{X},\linspan{Y}) &= 2\rank \mat{\hat{L} & \bm{r} - \bm{x} \\ 0 & \hat{E}} - {(\mu + \delta)}. \nonumber
  \end{align}
\end{theorem}
\begin{IEEEproof}
  See the Appendix.
\end{IEEEproof}
\medskip

A consequence of Theorem~\ref{thm:reduction-and-Delta} is that,
under the lifting construction, the decoding problem
(\ref{eq:RLNCC.decoding-rule}) for random network coding can be
abstracted to a generalized decoding problem for rank-metric codes.
More precisely, if we cascade an RLNCC, at the input, with a device
that takes $\bm{x}$ to its lifting $X=\mat{I & \bm{x}}$ and, at the output,
with a device that takes $Y$ to its reduction $(\bm{r},\hat{L},\hat{E})$,
then the decoding problem (\ref{eq:RLNCC.decoding-rule}) reduces
to the following problem:

\bigskip
\noindent\textbf{Generalized Decoding Problem for Rank-Metric Codes:}
{\slshape
Let $\mathcal{C} \subseteq \Fq^{n \times m}$ be a rank-metric code. Given a received tuple $(\bm{r},\hat{L},\hat{E}) \in \Fq^{n \times m} \times \Fq^{n \times \mu} \times \Fq^{\delta \times m}$ with $\rank \hat{L} = \mu$ and $\rank \hat{E} = \delta$, find
\begin{equation}\label{eq:rankmetric.generalized-decoding}
  \hat{\bm{x}} = \argmin_{\bm{x} \in \mathcal{C}}\, \rank \mat{\hat{L} & \bm{r} - \bm{x} \\ 0 & \hat{E}}.
\end{equation}
}
\medskip

The problem above will be referred to as the generalized decoding problem
for rank-metric codes, or generalized rank decoding for short. Note that
the conventional rank decoding problem (\ref{eq:rank-decoding.standard})
corresponds to the special case where $\mu = \delta = 0$.

The remainder of this paper is devoted to the study of the
generalized rank decoding problem and to its solution in the case of MRD codes.

\section{A Generalized Decoding Problem for Rank-Metric Codes}
\label{sec:generalized-decoding}

In this section, we develop a perspective on
the generalized rank decoding problem that will prove
useful to the understanding of the correction capability
of rank-metric codes, as well as to the formulation of
an efficient decoding algorithm.

\subsection{Error Locations and Error Values}
\label{sec:locations-values}

Let $\mathcal{C} \in \Fq^{n\times m}$ be a rank-metric code.
For a transmitted codeword $\bm{x}$ and a received word $\bm{r}$,
define $\bm{e} \triangleq \bm{r} - \bm{x}$ as the error word.

Note that if an error word $\bm{e}$ has rank $\tau$, then we can
write $\bm{e} = LE$ for some full-rank matrices $L \in \Fq^{n \times \tau}$
and $E \in \Fq^{\tau \times m}$, as in (\ref{eq:rank.equivalent-definition}).
Let $L_1,\ldots,L_\tau \in \Fq^n$ denote the columns of $L$ and let $E_1,\ldots,E_\tau \in \Fq^{1 \times m}$ denote the rows of $E$.
Then we can expand $\bm{e}$ as a summation of outer products
\begin{equation}\label{eq:error.expansion}
  \bm{e} = LE = \sum_{j=1}^\tau L_j E_j.
\end{equation}

We will now borrow some terminology from classical coding theory.
Recall that an error vector $e \in \Fq^n$ of Hamming weight $\tau$ can
be expanded uniquely as a sum of products
\[
e = \sum_{j=1}^\tau I_{i_j} e_j
\]
where $1\leq i_1 < \cdots < i_\tau \leq n$ and $e_1,\ldots,e_\tau \in \Fq$.
The index $i_j$ (or the unit vector $I_{i_j}$) specifies the
\emph{location} of the $j$th error, while
$e_j$ specifies the \emph{value} of the $j$th error.

Analogously, in the sum-of-outer-products expansion
(\ref{eq:error.expansion})
we will refer to $L_1,\ldots,L_\tau$ as the \defn{error locations}
and to $E_1,\ldots,E_\tau$ as the \defn{error values}.
The location $L_j$ (a column vector) indicates that,
for $i=1,\ldots,n$, the $j$th error value $E_j$ (a row vector)
occurred in row $i$ multiplied by the coefficient $L_{ij}$.
Of course, $L_{ij}=0$ means that the $j$th error value is not
present in row $i$.

Note that, in contrast to the classical case, the distinction between error locations and error values in the rank metric is merely a convention. If we prefer to think of errors as occurring on columns rather than rows, then the roles of $L_j$ and $E_j$ would be interchanged. The same observation will also apply to any concept derived from the interpretation of these quantities as error locations and error values.

It is important to mention that, in contrast with classical coding theory,
the expansion (\ref{eq:error.expansion}) is not unique, since
\begin{equation}\nonumber
  \bm{e} = LE = LT^{-1} TE
\end{equation}
for any nonsingular $T \in \Fq^{\tau \times \tau}$.
Thus, strictly speaking, $L_1,\ldots,L_\tau$ and $E_1,\ldots,E_\tau$ are
just one possible set of error locations/values describing
the error word $\bm{e}$.

\subsection{Erasures and Deviations}
\label{erasures-deviations}

We now reformulate the generalized rank decoding problem
in a way that facilitates its understanding and solution.

First, observe that the problem (\ref{eq:rankmetric.generalized-decoding})
is equivalent to
the problem of finding an error word $\hat{\bm{e}}$,
given by
\begin{equation}\label{eq:generalized-decoding.error-word}
  \hat{\bm{e}} = \argmin_{\bm{e} \in \bm{r} - \mathcal{C}}\, \rank \mat{\hat{L} & \bm{e} \\ 0 & \hat{E}},
\end{equation}
from which the output of the decoder can be computed
as $\hat{\bm{x}} = \bm{r} - \hat{\bm{e}}$.

\medskip
\begin{proposition}\label{prop:rank-expansion.equivalence}
  Let $\bm{e} \in \Fq^{n \times m}$, $\hat{L} \in \Fq^{n \times \mu}$ and $\hat{E} \in \Fq^{\delta \times n}$. The following statements are equivalent:
  \begin{enumerate}
    \item $\tau^* = \rank \mat{\hat{L} & \bm{e} \\ 0 & \hat{E}}.$
    \item $\tau^*-\mu-\delta$ is the minimum value of
    \begin{equation}\nonumber
      \rank(\bm{e} - \hat{L}E^{(1)} - L^{(2)}\hat{E})
    \end{equation} for all $E^{(1)} \in \Fq^{\mu \times m}$ and all $L^{(2)} \in \Fq^{n \times \delta}$.
    \item $\tau^*$ is the minimum value of $\tau$ for which there exist $L_1,\ldots,L_\tau \in \Fq^{n}$ and $E_1,\ldots,E_\tau \in \Fq^{1 \times m}$ satisfying:
  \begin{align}
    \bm{e} &= \sum_{j=1}^{\tau} L_j E_j  \nonumber \\
    L_j &= \hat{L}_j, \quad j=1,\ldots,\mu  \nonumber \\
    E_{\mu + j} &= \hat{E}_j, \quad j = 1,\ldots,\delta. \nonumber
  \end{align}

  \end{enumerate}
\end{proposition}
\begin{IEEEproof}
  See the Appendix.
\end{IEEEproof}
\medskip

With the help of Proposition~\ref{prop:rank-expansion.equivalence}, the influence of $\hat{L}$ and $\hat{E}$ in the decoding problem can be interpreted as follows. Suppose $\bm{e} \in \bm{r} - \mathcal{C}$ is the unique solution to (\ref{eq:generalized-decoding.error-word}). Then $\bm{e}$ can be expanded as $\bm{e} = \sum_{j=1}^\tau L_j E_j$, where $L_1,\ldots,L_\mu$ and $E_{\mu+1},\ldots,E_{\mu+\delta}$ are \emph{known} to the decoder. In other words, the decoding problem is facilitated, since the decoder has side information about the expansion of $\bm{e}$.

Recall the terminology of Section~\ref{sec:locations-values}.
Observe that, for $j \in \{1,\ldots,\mu\}$, the decoder knows
the \emph{location} of the $j$th error term but not its value,
while for $j \in \{\mu+1,\ldots,\mu+\delta\}$, the decoder knows
the \emph{value} of the $j$th error term but not its location.
Since in classical coding theory knowledge of an error location
but not its value corresponds to an erasure, we will adopt a similar
terminology here.  However we will need to introduce a new
term to handle the case where the value of an error
is known, but not its location.
In the expansion (\ref{eq:error.expansion}) of the error word,
each term $L_j E_j$ will be called
\begin{itemize}
\item an \defn{erasure}, if $L_j$ is known;
\item a \defn{deviation}, if $E_j$ is known; and
\item a \defn{full error} (or simply an \emph{error}),
if neither $L_j$ nor $E_j$ are known.
\end{itemize}
Collectively, erasures, deviations and errors will be referred
to as ``errata.''
We say that an errata pattern is \defn{correctable} when
(\ref{eq:rankmetric.generalized-decoding}) has a unique solution equal
to the original transmitted codeword.

The following theorem characterizes
the errata-correction capability of rank-metric codes.
\begin{theorem}\label{thm:rankmetric.correction-capability}
A rank-metric code $\mathcal{C} \subseteq \Fq^{n \times m}$
of minimum distance $d$ is able to correct every pattern of
$\epsilon$ errors, $\mu$ erasures and $\delta$ deviations
if and only if $2\epsilon + \mu + \delta \leq d-1$.
\end{theorem}
\begin{IEEEproof}
Let $\bm{x} \in \mathcal{C}$ be a transmitted codeword and let
$(\bm{r},\hat{L},\hat{E}) \in \Fq^{n \times m} \times \Fq^{n \times \mu} \times \Fq^{\delta \times m}$
be a received tuple such that
$\rank \mat{\hat{L} & \bm{r} - \bm{x}\\ 0 & \hat{E}}=\mu + \delta + \epsilon$.
Suppose $\bm{x}' \in \mathcal{C}$ is another codeword such that
$\rank \mat{\hat{L} & \bm{r} - \bm{x}'\\0 & \hat{E}}=\mu + \delta + \epsilon'$,
where $\epsilon' \leq \epsilon$.
From Proposition~\ref{prop:rank-expansion.equivalence}, we can write
  \begin{align}
    \bm{e}  = \bm{r} - \bm{x} &= \hat{L} E^{(1)} + L^{(2)} \hat{E} + L^{(3)} E^{(3)}  \nonumber \\
    \bm{e}' = \bm{r} - \bm{x}' &= \hat{L} E^{(4)} + L^{(5)} \hat{E} + L^{(6)} E^{(6)} \nonumber
  \end{align}
for some $E^{(1)},L^{(2)},\ldots,E^{(6)}$ with appropriate dimensions
such that $\rank L^{(3)} E^{(3)} = \epsilon$ and $\rank L^{(6)} E^{(6)} = \epsilon'$.

Thus,
\begin{equation}\nonumber
    \bm{e}-\bm{e}' = \hat{L} (E^{(1)}-E^{(4)}) + (L^{(2)}-L^{(5)}) \hat{E} + L^{(3)} E^{(3)} + L^{(6)} E^{(6)}
  \end{equation}
  and
  \begin{equation}\nonumber
    \rank(\bm{x}' - \bm{x}) = \rank (\bm{e}-\bm{e}') \leq \mu + \delta + \epsilon + \epsilon' \leq d-1
  \end{equation}
contradicting the minimum distance of the code.

Conversely, let $\bm{x},\bm{x}' \in \mathcal{C}$ be two codewords such that $\rank(\bm{x}' - \bm{x}) = d$. For all $\mu$, $\delta$ and $\epsilon$ such that $\mu + \delta + 2\epsilon \geq d$, we can write
  \begin{equation}\nonumber
    \bm{x}'-\bm{x} = L^{(1)} E^{(1)}+ L^{(2)} E^{(2)}+ L^{(3)} E^{(3)} + L^{(4)} E^{(4)}
  \end{equation}
where the four terms above have inner dimensions equal to $\mu$, $\delta$, $\epsilon$ and $\epsilon' = d-\mu-\delta-\epsilon$, respectively.
Let
  \begin{align}
    \bm{e}  &= L^{(1)} E^{(1)}+ L^{(2)} E^{(2)}+ L^{(3)} E^{(3)}  \nonumber \\
    \bm{e}' &= - L^{(4)} E^{(4)} \nonumber
  \end{align}
  and observe that $\bm{x}' - \bm{x} = \bm{e} - \bm{e}'$. Let $\bm{r} = \bm{x} + \bm{e} = \bm{x}' + \bm{e}'$, $\hat{L} = L^{(1)}$ and $\hat{E} = E^{(2)}$. Suppose that $\bm{x}$ is transmitted and the tuple $(\bm{r},\hat{L},\hat{E})$ is received. Then
  \begin{align}
    \rank \mat{\hat{L} & \bm{r} - \bm{x} \\ 0 & \hat{E}} &= \rank \mat{\hat{L} & \bm{e} \\ 0 & \hat{E}} = \mu + \delta + \epsilon  \nonumber \\
    \rank \mat{\hat{L} & \bm{r} - \bm{x}' \\ 0 & \hat{E}} &= \rank \mat{\hat{L} & \bm{e}' \\ 0 & \hat{E}} = \mu + \delta + \epsilon'. \nonumber
  \end{align}
Since $\epsilon' = d-\mu-\delta-\epsilon \leq \epsilon$,
it follows that $\bm{x}$ cannot be the unique solution to
(\ref{eq:rankmetric.generalized-decoding}) and therefore the
errata pattern cannot be corrected.
\end{IEEEproof}

\medskip

Theorem~\ref{thm:rankmetric.correction-capability} shows that,
similarly to erasures in the Hamming metric, erasures and deviations
cost half of an error in the rank metric.

Theorem~\ref{thm:rankmetric.correction-capability} also
shows that taking into account information about erasures and deviations
(when they occur) can strictly increase the error correction capability
of a rank-metric code. Indeed, suppose that an error word of
rank $t = \mu + \delta + \epsilon$ is applied to a codeword,
where $\mu$, $\delta$ and $\epsilon$ are the number of erasures,
deviations and full errors, respectively, in the errata pattern.
It follows that a conventional rank decoder (which ignores the
information about erasures and deviations) can only guarantee
successful decoding if $2t \leq d-1$, where $d$ is the minimum rank
distance of the code. On the other hand, a generalized rank decoder
requires only $2\epsilon + \mu + \delta \leq d-1$,
or $2t \leq d-1 + \mu + \delta$, in order to guarantee successful decoding.
In this case, the error correction capability is increased
by $(\mu + \delta)/2$ if a generalized rank decoder is used instead
of a conventional one.
\medskip

We conclude this section by comparing our generalized decoding problem
with previous decoding problems proposed for rank-metric codes.

There has been a significant amount of research on the problem of
correcting rank errors in the presence of ``row and column erasures''
\cite{Gabidulin++1992,Gabidulin.Pilipchuk2003,Richter.Plass2004:BerlekampMassey,Richter.Plass2004:ColumnErasures,Pilipchuk.Gabidulin2007},
where a row erasure means that all entries of that row are replaced
by an erasure symbol, and similarly for a column erasure.
The decoding problem in this setting is naturally defined as finding
a codeword such that, when the erased entries in the received word are
replaced by those of the codeword, the difference between this new matrix and the codeword
has the smallest possible rank.
We now show that this problem is a special case of (\ref{eq:rankmetric.generalized-decoding}).

First, we force the received word $\bm{r}$ to be in $\Fq^{n \times m}$
by replacing each erasure symbol with an arbitrary symbol in $\Fq$, say $0$.
Suppose that the rows $i_1,\ldots,i_\mu$ and the columns $k_1,\ldots,k_\delta$
have been erased. Let $\hat{L} \in \Fq^{n \times \mu}$ be given by
$\hat{L}_{i_j,j} = 1$ and $\hat{L}_{i,j} = 0,\, \forall i \neq i_j$,
for $j=1,\ldots,\mu$ and let $\hat{E} \in \Fq^{\delta \times m}$ be given
by $\hat{E}_{j,k_j} = 1$ and $\hat{E}_{j,k} = 0,\, \forall k \neq k_j$,
for $j=1,\ldots,\delta$.
Since
\begin{equation}\label{eq:row-column-erasures.received-matrix}
  \mat{\hat{L} & \bm{r} - \bm{x} \\ 0 & \hat{E}}
  =
  \mat{\hat{L} & \bm{r} \\ 0 & \hat{E}}
  -
  \mat{0 & \bm{x} \\ 0 & 0}
\end{equation}
it is easy to see that we can perform column operations on
(\ref{eq:row-column-erasures.received-matrix}) to replace the
erased rows of $\bm{r}$ with the same entries as $\bm{x}$, and
similarly we can perform row operations on
(\ref{eq:row-column-erasures.received-matrix}) to replace the
erased columns of $\bm{r}$ with the same entries as $\bm{x}$.
The decoding problem (\ref{eq:rankmetric.generalized-decoding})
is unchanged by these operations and reduces exactly to the decoding
problem with ``row and column erasures'' described in the previous paragraph. An example is given
below.

\medskip
\begin{example}
  Let $n=m=3$. Suppose the third row and the second column have been erased in the received word. Then
  \begin{equation}\nonumber
    \bm{r} = \mat{
      r_{11} & 0 & r_{13} \\
      r_{21} & 0 & r_{23} \\
      0      & 0 & 0
   },
    \quad \hat{L} = \mat{0 \\ 0 \\ 1},
    \quad \hat{E} = \mat{0 & 1 & 0}.
  \end{equation}
  Since
  \begin{equation}\nonumber
    \mat{
      0 & r_{11} & 0 & r_{13} \\
      0 & r_{21} & 0 & r_{23} \\
      1 & 0      & 0 & 0      \\
      0 & 0      & 1 & 0
   }
    \quad \text{and} \quad
    \mat{
      0 & r_{11} & x_{12} & r_{13} \\
      0 & r_{21} & x_{22} & r_{23} \\
      1 & x_{31} & x_{32} & x_{33} \\
      0 & 0      & 1      & 0
   }
  \end{equation}
  are row equivalent, we obtain that
  \begin{align}
    \rank \mat{\hat{L} & \bm{r} - \bm{x} \\ 0 & \hat{E}}
    &= \rank \mat{
      0 & r_{11}-x_{11} & 0 & r_{13}-x_{13} \\
      0 & r_{21}-x_{21} & 0 & r_{23}-x_{23} \\
      1 & 0             & 0 & 0             \\
      0 & 0             & 1 & 0
   }  \nonumber \\
    &= 2 + \rank \mat{
      r_{11}-x_{11} & 0 & r_{13}-x_{13} \\
      r_{21}-x_{21} & 0 & r_{23}-x_{23} \\
      0             & 0 & 0
   } \nonumber
  \end{align}
which is essentially the same objective function as in the
decoding problem with ``row and column erasures'' described above.
\qed
\end{example}
\medskip

While row/column erasures are a special case of erasures/deviations, it also true that the latter can always be transformed into the former. This can be accomplished by multiplying all rank-metric codewords to the left and to the right by nonsingular matrices in such a way that the corresponding $\hat{L}_j$ and $\hat{E}_j$ become unit vectors. The drawback of this approach, as pointed out in Section~\ref{sec:decoding-lifted-codes}, is that the structure of the code is changed at each decoding instance, which may raise complexity and/or implementation issues. Thus, it is practically more advantageous to fix the structure of the code and construct a decoder that can handle the generalized notions of erasures and deviations. This is the approach we take in the next section.

\section{Decoding Gabidulin Codes with Errors, Erasures and Deviations}
\label{sec:decoding-error-erasures-deviations}

In this section, we turn our attention to the design of an efficient rank
decoder that can correct any pattern of $\epsilon$ errors, $\mu$ erasures
and $\delta$ deviations satisfying $2 \epsilon + \mu + \delta \leq d-1$,
where $d$ is the minimum rank distance of the code. Our decoder is applicable
to Gabidulin codes, a class of MRD codes proposed in \cite{Gabidulin1985}.

\subsection{Preliminaries}

Rank-metric codes in $\Fq^{n \times m}$ are typically constructed as
block codes of length $n$ over the extension field $\Fqm$. More precisely,
by fixing a basis for $\Fqm$ as an $m$-dimensional vector space over $\Fq$,
we can regard any element of $\Fqm$ as a \emph{row} vector of length $m$
over $\Fq$ (and vice-versa). Similarly, we can regard any \emph{column}
vector of length $n$ over $\Fqm$ as an $n \times m$ matrix over $\Fq$
(and vice-versa). All concepts previously defined for matrices in
$\Fq^{n \times m}$ can be naturally applied to vectors in $\Fqm^n$;
in particular, the rank of a vector $\bm{x} \in \Fqm^n$ is the rank
of $\bm{x}$ as an $n \times m$ matrix over~$\Fq$.

\subsubsection{Gabidulin Codes}
\label{sec:gabidulin}

In order to simplify notation, let $[i]$ denote $q^i$. A Gabidulin code is a linear $(n,k)$ code over $\Fqm$ defined by the parity-check matrix
\begin{equation}\nonumber
H = \mat{
  h_1^{[0]} & h_2^{[0]} & \cdots & h_n^{[0]} \\
  h_1^{[1]} & h_2^{[1]} & \cdots & h_n^{[1]} \\
  \vdots & \vdots  & \ddots & \vdots \\
  h_1^{[n-k-1]} & h_2^{[n-k-1]} & \cdots & h_n^{[n-k-1]}
  }
\end{equation}
%$H = [H_{ij}]$, $H_{ij} = h_{j}^{[i-1]}$, $1\leq i \leq n-k$, $1\leq j \leq n$,
where the elements $h_1,\ldots,h_n \in \Fqm$ are linearly independent
over $\Fq$ (note that $n \leq m$ is required).
The minimum rank distance of a Gabidulin code is $d = n-k+1$,
satisfying the Singleton bound in the rank metric \cite{Gabidulin1985}.

\subsubsection{Linearized Polynomials}
\label{sec:linearized}

A class of polynomials that play an important role in the study
of rank-metric codes are the \defn{linearized polynomials}
\cite[Sec.~3.4]{Lidl.Niederreiter}.
A linearized polynomial (or $q$-polynomial) over $\Fqm$ is a
polynomial of the form
\begin{equation}\nonumber
 f(x) = \sum_{i=0}^{t} f_i x^{[i]}
\end{equation}
where $f_i \in \Fqm$. If $f_t \neq 0$, we call
$t$ the $q$-degree of $f(x)$. Linearized polynomials receive
their name because of the following property: for any $a_1,a_2 \in \Fq$ and any $\beta_1,\beta_2 \in \Fqm$,
\[
f(a_1 \beta_1 + a_2 \beta_2) = a_1 f(\beta_1) + a_2 f(\beta_2).
\]
That is, evaluation of a linearized polynomial is a map $\Fqm \to \Fqm$
that is linear over $\Fq$. In particular, the set of all roots in $\Fqm$
of a linearized polynomial is a subspace of $\Fqm$.

Let $A(x)$ and $B(x)$ be linearized polynomials of
$q$-degrees $t_A$ and $t_B$, respectively.
The symbolic product of $A(x)$ and $B(x)$ is defined
as the polynomial $A(x) \otimes B(x) \triangleq A(B(x))$.
It is easy to verify that $P(x) = A(x) \otimes B(x)$ is a
linearized polynomial of $q$-degree $t = t_A + t_B$ whose
coefficients can be computed as
\begin{equation}\nonumber
  P_\ell = \sum_{i = \max\{0,\ell-t_B\}}^{\min\{\ell,t_A\}} A_i B_{\ell-i}^{[i]} = \sum_{j = \max\{0,\ell-t_A\}}^{\min\{\ell,t_B\}} A_{\ell-j} B_{j}^{[\ell-j]}
\end{equation}
for $\ell=0,\ldots,t$.
In particular, if $t_A \leq t_B$, then
\begin{equation}\label{eq:lin-poly.product.small-left}
  P_\ell = \sum_{i = 0}^{t_A} A_i B_{\ell-i}^{[i]}, \qquad t_A \leq \ell \leq t_B,
\end{equation}
while if $t_B \leq t_A$, then
\begin{equation}\label{eq:lin-poly.product.small-right}
  P_\ell = \sum_{j = 0}^{t_B} A_{\ell-j} B_{j}^{[\ell-j]}, \qquad t_B \leq \ell \leq t_A.
\end{equation}
It is known that the set of linearized polynomials over $\Fqm$ together
with the operations of polynomial addition and symbolic multiplication
forms a noncommutative ring with identity having many of the properties
of a Euclidean domain.

We define the \defn{$q$-reverse} of a linearized polynomial
$f(x) = \sum_{i=0}^t f_i x^{[i]}$ as the polynomial
$\bar{f}(x) = \sum_{i=0}^t \bar{f}_i x^{[i]}$ given by
$\bar{f}_i = f_{t-i}^{[i-t]}$ for $i=0,\ldots,t$.
(When $t$ is not specified we will assume that $t$ is the
$q$-degree of $f(x)$.)

For a set $\mathcal{S} \subseteq \Fqm$, define the
\defn{minimal linearized polynomial} of $\mathcal{S}$
(with respect to $\Fqm$), denoted $M_\mathcal{S}(x)$ or
$\minpoly\{\mathcal{S}\}(x)$, as the monic linearized polynomial
over $\Fqm$ of least degree whose root space contains $\mathcal{S}$.
It can be shown that $M_\mathcal{S}(x)$ is given by
\begin{equation}\nonumber
  M_\mathcal{S}(x) \triangleq \prod_{\beta \in \linspan{\mathcal{S}}} (x - \beta)
\end{equation}
so the $q$-degree of $M_\mathcal{S}(x)$ is equal to
$\dim \linspan{\mathcal{S}}$.
Moreover, if $f(x)$ is any linearized polynomial whose root space
contains $\mathcal{S}$, then
\begin{equation}\nonumber
  f(x) = Q(x) \otimes M_\mathcal{S}(x)
\end{equation}
for some linearized polynomial $Q(x)$.
This implies that
$M_{\mathcal{S} \cup \{\alpha\}}(x) = M_{M_{\mathcal{S}}(\alpha)}(x) \otimes M_\mathcal{S}(x)$ for any $\alpha$.
Thus, $M_\mathcal{S}(x)$ can be computed in $O(t^2)$ operations
in $\Fqm$ by taking a basis $\{\alpha_1,\ldots,\alpha_t\}$ for
$\linspan{S}$ and computing $M_{\{\alpha_1,\ldots,\alpha_i\}}(x)$
recursively for $i=1,\ldots,t$.

\subsubsection{Decoding of Gabidulin Codes}
\label{sec:decoding-gabidulin}

Recall that, in the conventional rank decoding problem with $\tau$ errors,
where $2\tau \leq d-1$, we are given a received word $\bm{r} \in \Fqm^n$
and we want to find the unique error word $\bm{e} \in \bm{r} - \mathcal{C}$
such that $\rank \bm{e} = \tau$. We review below the usual decoding procedure,
which consists of finding error values $E_1,\ldots,E_\tau \in \Fqm$
and error locations $L_1,\ldots,L_\tau \in \Fq^n$ such that
$\bm{e} = \sum_{j=1}^\tau L_j E_j$.

Since $\bm{e} \in \bm{r} - \mathcal{C}$, we can form the \defn{syndromes}
\begin{equation}\nonumber%\label{eq:syndromes}
  [S_0,\ldots,S_{d-2}]^T \triangleq H \bm{r} = H \bm{e}
\end{equation}
which can then be related to the error values and error locations according to
\begin{align}
  S_\ell &= \sum_{i=1}^n h_i^{[\ell]} e_i = \sum_{i=1}^n h_i^{[\ell]} \sum_{j=1}^\tau L_{ij} E_j \nonumber \\
  \label{eq:syndrome-equation}
  &= \sum_{j=1}^\tau X_j^{[\ell]} E_j ,\quad \ell = 0,\ldots,d-2
\end{align}
where
\begin{equation}\label{eq:locators}
  X_j = \sum_{i=1}^n L_{ij} h_i,\quad j = 1,\ldots,\tau
\end{equation}
are called the \defn{error locators} associated with $L_1,\ldots,L_\tau$.

Suppose, for now, that the error values $E_1,\ldots,E_\tau$
(which are essentially $\tau$ linearly independent elements
satisfying $\linspan{\bm{e}} = \linspan{E_1,\ldots,E_\tau}$)
have already been determined. Then the error locators can be
determined by solving (\ref{eq:syndrome-equation}) or, equivalently,
by solving
\begin{equation} \label{eq:syndrome-equation.inverted}
  \bar{S}_{\ell} = S_{d-2-\ell}^{[\ell-d+2]} = \sum_{j=1}^\tau E_j^{[\ell-d+2]} X_j ,\quad \ell = 0,\ldots,d-2
\end{equation}
which is a system of equations of the form
\begin{equation} \label{eq:gab-alg.equation}
  B_{\ell} = \sum_{j=1}^\tau A_j^{[\ell]} X_j  ,\quad \ell = 0,\ldots,d-2
\end{equation}
consisting of $d-1$ linear equations (over $\Fqm$) in
$\tau$ unknowns $X_1,\ldots,X_\tau$.
Such a system is known to have a unique solution (whenever one exists)
provided $\tau \leq d-1$ and $A_1,\ldots,A_{\tau}$ are
linearly independent (see \cite{MacWilliams.Sloane,Lidl.Niederreiter}).
Moreover, a solution to (\ref{eq:gab-alg.equation}) can be found efficiently
in $O(d^2)$ operations in $\Fqm$ by an algorithm proposed by
Gabidulin \cite[pp.~9--10]{Gabidulin1985}.

After the error locators have been found, the error locations
$L_1,\ldots,L_\tau$ can be easily recovered by solving (\ref{eq:locators}).
More precisely, let $\bm{h} \in \Fq^{n \times m}$ be the matrix whose
rows are $h_1,\ldots,h_n$, and let $Q \in \Fq^{m \times n}$ be a right
inverse of $\bm{h}$, i.e., $\bm{h}Q = I_{n \times n}$.
Then
\begin{equation}\nonumber
  L_{ij} = \sum_{k=1}^{m} X_{jk} Q_{ki}, \quad i=1,\ldots,n, \quad j=1,\ldots,\tau.
\end{equation}

The computation of error values can be done indirectly via an
\defn{error span polynomial} $\sigma(x)$.
Let $\sigma(x)$ be a linearized polynomial of $q$-degree $\tau$ having
as roots all linear combinations of $E_1,\ldots,E_\tau$.
Then, $\sigma(x)$ can be related to the \defn{syndrome polynomial}
\begin{equation}\nonumber
  S(x) = \sum_{j=0}^{d-2} S_j x^{[j]}
\end{equation}
through the following \defn{key equation}:
\begin{equation}\label{eq:original-key-equation}
  \sigma(x) \otimes S(x) \equiv \omega(x) \mod x^{[d-1]}
\end{equation}
where $\omega(x)$ is a linearized polynomial of $q$-degree $\leq \tau-1$.

An equivalent way to express (\ref{eq:original-key-equation}) is
\begin{equation}\label{eq:original-key-equation.alternative}
  \sum_{i=0}^{\tau} \sigma_{i} S_{\ell-i}^{[i]} = 0, \quad \ell = \tau,\ldots, d-2.
\end{equation}

This key equation can be efficiently solved in $O(d^2)$ operations
in $\Fqm$ by the modified Berlekamp-Massey algorithm proposed
in \cite{Richter.Plass2004:BerlekampMassey}, provided $2\tau \leq d-1$.

After the error span polynomial is found, the error values can be
obtained by computing a basis $E_1,\ldots,E_\tau$ for the root
space of $\sigma(x)$. This can be done either by the probabilistic algorithm
in \cite{Skachek.Roth2008}, in an average of $O(dm)$ operations in
$\Fqm$, or by the methods in \cite{Berlekamp}, which take at most
$O(m^3)$ operations in $\Fq$ plus $O(dm)$ operations in $\Fqm$.

\subsection{A Modified Key Equation Incorporating Erasures and Deviations}
\label{sec:modified-key-equation}

In the general rank decoding problem with $\epsilon$ errors, $\mu$ erasures
and $\delta$ deviations, where $2\epsilon + \mu + \delta \leq d-1$,
we are given a received tuple
$(\bm{r}, \hat{L}, \hat{E}) \in \Fqm^n \times \Fq^{n \times \mu} \times \Fqm^\delta$
and we want to find the unique error word
$\bm{e} \in \bm{r} - \mathcal{C}$ such that
$\rank \mat{\hat{L} & \bm{e} \\ 0 & \hat{E}} = \epsilon + \mu + \delta \triangleq \tau$ (along with the value of $\epsilon$, which is not known a priori).

First, note that if we can find a linearized polynomial
$\sigma(x)$ of $q$-degree at most $\tau \leq d-1$ satisfying
$\sigma(e_i) = 0$, $i=1,\ldots,n$, then the error word can be
determined in the same manner as in Section~\ref{sec:decoding-gabidulin}.

According to Proposition~\ref{prop:rank-expansion.equivalence},
we can write the error word as $\bm{e} = \sum_{j=1}^\tau L_j E_j$
for some $L_1,\ldots,L_\tau \in \Fq^n$ and $E_1,\ldots,E_\tau \in \Fqm$
satisfying $L_j = \hat{L}_j$, $j=1,\ldots,\mu$,
and $E_{\mu+j} = \hat{E}_j$, $j=1,\ldots,\delta$.
Let $\sigma_D(x)$, $\sigma_F(x)$ and $\sigma_U(x)$
be linearized polynomials of smallest $q$-degrees satisfying
\begin{align}
 \sigma_D(E_j) &= 0, \quad j=\mu+1,\ldots,\mu+\delta  \nonumber \\
 \sigma_F(\sigma_D(E_j)) &= 0, \quad j=\mu+\delta+1,\ldots,\tau  \nonumber \\
 \sigma_U(\sigma_F(\sigma_D(E_j))) &= 0, \quad j=1,\ldots,\mu. \nonumber
\end{align}
Clearly, the $q$-degrees of $\sigma_D(x)$ and $\sigma_F(x)$
are $\delta$ and $\epsilon$, respectively, and the $q$-degree of $\sigma_U(x)$ is at most $\mu$.

Define the \defn{error span polynomial}
\begin{equation}\nonumber%\label{eq:error-span-polynomial}
  \sigma(x) = \sigma_U(x) \otimes \sigma_F(x) \otimes \sigma_D(x).
\end{equation}
Then $\sigma(x)$ is a linearized polynomial of $q$-degree $\leq \tau$ satisfying
\begin{equation}\nonumber
  \sigma(e_i) = \sigma(\sum_{j=1}^\tau L_{ij} E_j) = \sum_{j=1}^\tau L_{ij} \sigma(E_j) = 0, \quad i=1,\ldots,n.
\end{equation}
Thus, since $\sigma_D(x)$ can be readily determined from
$\hat{E}$, decoding reduces to the determination of
$\sigma_F(x)$ and $\sigma_U(x)$.

Now, let $\lambda_U(x)$ be a linearized polynomial of $q$-degree
$\mu$ satisfying
\begin{equation}\nonumber
  \lambda_U(X_j) = 0, \quad j=1,\ldots,\mu
\end{equation}
and let $\bar{\lambda}_U(x)$ be the $q$-reverse of $\lambda_U(x)$.
We define an \defn{auxiliary syndrome polynomial} as
\begin{equation}\nonumber
  S_{DU}(x) = \sigma_D(x) \otimes S(x) \otimes \bar{\lambda}_U(x).
\end{equation}
Observe that $S_{DU}(x)$ incorporates all the information that is known
at the decoder, including erasures and deviations.

Our modified key equation is given in the following theorem.
\begin{theorem} \label{thm:key-equation}
\begin{equation}\label{eq:key-equation}
  \sigma_F(x) \otimes S_{DU}(x) \equiv \omega(x) \mod x^{[d-1]}
\end{equation}
where $\omega(x)$ is a linearized polynomial of $q$-degree $\leq \tau-1$.
\end{theorem}
\begin{IEEEproof}
Let $\omega(x) = \sigma_F(x) \otimes S_{DU}(x) \bmod x^{[d-1]}$.
If $\tau \geq d-1$, we have nothing to prove, so let us assume $\tau \leq d-2$.
We will show that $\omega_\ell = 0$ for $\ell = \tau,\ldots,d-2$.

Let $\sigma_{FD}(x) = \sigma_F(x) \otimes \sigma_D(x)$ and $S_{FD}(x) = \sigma_{FD}(x) \otimes S(x)$.
According to (\ref{eq:lin-poly.product.small-left}), for
$\epsilon + \delta \leq \ell \leq d-2$ we have
\begin{align}
  S_{FD,\ell} &= \sum_{i=0}^{\epsilon+\delta} \sigma_{FD,i} S_{\ell-i}^{[i]} = \sum_{i=0}^{\epsilon+\delta}
\sigma_{FD,i} \left(\sum_{j=1}^\tau X_j^{[\ell-i]} E_j \right)^{[i]} \nonumber \\
  \label{eq:sigma-U-roots}
  &= \sum_{j=1}^\tau X_j^{[\ell]} \sigma_{FD}(E_j) = \sum_{j=1}^\mu X_j^{[\ell]} \beta_j,
\end{align}
where
\begin{equation}\nonumber
  \beta_j = \sigma_{FD}(E_j), \quad j=1,\ldots,\mu.
\end{equation}

Note that
$\sigma_F(x) \otimes S_{DU}(x) = S_{FD}(x) \otimes \bar{\lambda}_U(x)$.
Using (\ref{eq:lin-poly.product.small-right}) and (\ref{eq:sigma-U-roots}),
for $\mu+ \epsilon + \delta \leq \ell \leq d-2$ we have
\begin{align}
  \omega_\ell &= \sum_{i=0}^{\mu} \bar{\lambda}_{U,i}^{[\ell-i]} S_{FD,\ell-i} = \sum_{i=0}^{\mu} \lambda_{U,\mu-i}^{[\ell-\mu]} \sum_{j=1}^\mu X_j^{[\ell-i]} \beta_j \nonumber \\
  &= \sum_{j=1}^\mu \sum_{i=0}^{\mu} \lambda_{U,i}^{[\ell-\mu]} X_j^{[\ell-\mu+i]} \beta_j = \sum_{j=1}^\mu \lambda_{U}(X_j)^{[\ell-\mu]} \beta_j = 0. \nonumber
\end{align}
This completes the proof of the theorem.
\end{IEEEproof}

The key equation can be equivalently expressed as
\begin{equation}\label{eq:key-equation.alternative}
  \sum_{i=0}^{\epsilon} \sigma_{F,i} S_{DU,\ell-i}^{[i]} = 0, \quad \ell = \mu+\delta+\epsilon,\ldots, d-2.
\end{equation}
Note that this key equation reduces to the original key equation
(\ref{eq:original-key-equation}) when there are no erasures or deviations.
Moreover, it can be solved by the same methods as the original key equation
(\ref{eq:original-key-equation}), e.g., using the Euclidean algorithm
for linearized polynomials \cite{Gabidulin1985} or using the modified
Berlekamp-Massey algorithm from \cite{Richter.Plass2004:BerlekampMassey},
provided $2\epsilon \leq d-1 -\mu -\delta$ (which is true by assumption). Note that a small adjustment needs to be made so that (\ref{eq:key-equation.alternative}) becomes indeed equivalent to (\ref{eq:original-key-equation.alternative}); namely, we should choose $S_\ell$ in (\ref{eq:original-key-equation.alternative}) as $S_{\ell} = S_{DU,\ell+\mu+\delta}$ and replace $d$ with $d-\mu-\delta$.

After computing $\sigma_F(x)$, we still need to determine $\sigma_U(x)$.
In the proof of Theorem~\ref{thm:key-equation}, observe that
(\ref{eq:sigma-U-roots}) has the same form as (\ref{eq:gab-alg.equation});
thus, $\beta_1,\ldots,\beta_\mu$ can be computed using Gabidulin's algorithm \cite[pp.~9--10]{Gabidulin1985},
since $S_{FD}(x)$ and $X_1,\ldots,X_\mu$ are known.
Finally, $\sigma_U(x)$ can be obtained as
$\sigma_U(x) = \minpoly\{\beta_1,\ldots,\beta_\mu\}$.

\subsection{Summary of the Algorithm and Complexity Analysis}
\label{sec:summary-complexity}

The complete algorithm for decoding Gabidulin codes with erasures
and deviations is summarized in Fig.~\ref{fig:dec-alg1}.
\begin{figure}
\hrulefill
\begin{center}
\begin{minipage}{1\columnwidth}
  \textbf{Input:} received tuple $(\bm{r}, \hat{L}, \hat{E}) \in \Fqm^n \times \Fq^{n \times \mu} \times \Fqm^\delta$. \\
  \textbf{Output:} error word $\bm{e} \in \Fqm^n$.
  \begin{enumerate}
  \item \emph{Computing the auxiliary syndrome polynomial}:\\ Compute
  \begin{enumerate}
    \item \label{item:syndromes} $S_\ell = \sum_{i=1}^n h_i^{[\ell]} r_i$, $\ell = 0,\ldots,d-2$
    \item \label{item:locations-locators} $\hat{X}_j = \sum_{i=1}^{n} \hat{L}_{ij} h_i$, $j=1,\ldots,\mu$
    \item \label{item:minpoly-erasures} $\lambda_U(x) = \minpoly\{\hat{X}_1,\ldots,\hat{X}_\mu\}$
    \item \label{item:minpoly-deviations} $\sigma_D(x) = \minpoly\{\hat{E}_1,\ldots,\hat{E}_\delta\}$, and
    \item \label{item:mult-SDU} $S_{DU}(x) = \sigma_D(x) \otimes S(x) \otimes \bar{\lambda}_U(x)$.
  \end{enumerate}
  \item \emph{Computing the error span polynomial}:
  \begin{enumerate}
    \item \label{item:key-equation} Use the Berlekamp-Massey algorithm \cite{Richter.Plass2004:BerlekampMassey} to find $\sigma_F(x)$ that solve the key equation (\ref{eq:key-equation}).
    \item \label{item:mult-SED} Compute $S_{FD}(x) = \sigma_{F}(x) \otimes \sigma_{D}(x) \otimes S(x)$.
    \item \label{item:gab-alg-intermediate} Use Gabidulin's algorithm \cite{Gabidulin1985} to find $\beta_1,\ldots,\beta_\mu \in \Fqm$ that solve (\ref{eq:sigma-U-roots}).
    \item \label{item:minpoly-intermediate} Compute $\sigma_U(x) = \minpoly\{\beta_1,\ldots,\beta_\mu\}$ and
    \item \label{item:mult-UED} $\sigma(x) = \sigma_U(x) \otimes \sigma_{F}(x) \otimes \sigma_{D}(x)$.
  \end{enumerate}
  \item \emph{Finding the roots of the error span polynomial}:\\
    \label{item:find-roots} Use either the algorithm in \cite{Skachek.Roth2008} or the methods in \cite{Berlekamp} to find a basis $E_1,\ldots,E_\tau \in \Fqm$ for the root space of $\sigma(x)$.
  \item \emph{Finding the error locations}:
  \begin{enumerate}
    \item \label{item:gab-alg-locators} Solve (\ref{eq:syndrome-equation.inverted}) using Gabidulin's algorithm \cite{Gabidulin1985} to find the error locators $X_1,\ldots,X_\tau \in \Fqm$.
    \item \label{item:locators-locations} Compute the error locations $L_{ij} = \sum_{k=1}^{m} X_{jk} Q_{ki}$, $i=1,\ldots,n$, $j=1,\ldots,\tau$.
    \item \label{item:error-word} Compute the error word $\bm{e} = \sum_{j=1}^\tau L_j E_j$.
  \end{enumerate}
  \end{enumerate}
\end{minipage}
\end{center}
\hrulefill
  \caption{Generalized decoding algorithm for Gabidulin codes.} \label{fig:dec-alg1}
\end{figure}
We now estimate the complexity of this algorithm.

Steps~\ref{item:mult-SDU}), \ref{item:mult-SED}) and \ref{item:mult-UED})
are symbolic multiplications of linearized polynomials and can be performed
in $O(d^2)$ operations in $\Fqm$.
Steps~\ref{item:minpoly-erasures}), \ref{item:minpoly-deviations}) and
\ref{item:minpoly-intermediate}) involve finding a minimal linearized
polynomial, which takes $O(d^2)$ operations in $\Fqm$.
Steps~\ref{item:locations-locators}), \ref{item:locators-locations}) and
\ref{item:error-word}) are matrix multiplications and take $O(dnm)$
operations in $\Fq$ only.
Both instances \ref{item:gab-alg-intermediate}) and
\ref{item:gab-alg-locators}) of Gabidulin's algorithm and also
the Berlekamp-Massey algorithm in step~\ref{item:key-equation})
take $O(d^2)$ operations in $\Fqm$.

The most computationally demanding steps are \ref{item:syndromes}) computing the syndromes and \ref{item:find-roots}) finding a basis for
the root space of the error span polynomial. The former can be implemented
in a straightforward manner using $O(dn)$ operations in $\Fqm$,
while the latter can be performed using an average of $O(dm)$ operations
in $\Fqm$ with the algorithm in \cite{Skachek.Roth2008} (although the method
described in \cite{Berlekamp} will usually perform faster when $m$ is small).

We conclude that the overall complexity of the algorithm is $O(d m)$
operations in $\Fqm$.

\subsection{An Equivalent Formulation Based on the Error Locator Polynomial}
\label{sec:equivalent-error-locator}

Due to the perfect duality between error values and error locators
(both are elements of $\Fqm$), it is also possible to derive
a decoding algorithm based on an error \emph{locator} polynomial
that contains all the error locators as roots.

Let the auxiliary syndrome polynomial be defined as
\begin{equation}\nonumber
  S_{UD}(x) = \lambda_U(x) \otimes \bar{S}(x) \otimes \bar{\sigma}_D(x^{[d-2]})^{[-d+2]}
\end{equation}
where $\bar{\sigma}_D(x)$ is the $q$-reverse of $\sigma_D(x)$ and $\bar{S}(x)$
is the $q$-reverse of $S(x)$.

Let $\lambda_F(x)$ be a linearized polynomial of $q$-degree $\epsilon$
such that $\lambda_F(\lambda_U(X_i)) = 0$, for $i =\mu+\delta+1,\ldots,\tau$.
We have the following key equation:
\begin{theorem} \label{thm:key-equation2}
\begin{equation}\label{eq:key-equation2}
  \lambda_F(x) \otimes S_{UD}(x) \equiv \psi(x) \mod x^{[d-1]}
\end{equation}
where $\psi(x)$ is a linearized polynomial of $q$-degree $\leq \tau-1$.
\end{theorem}
\begin{IEEEproof}
The proof is similar to that of Theorem~\ref{thm:key-equation} and will be omitted.
%
%Let $\psi(x) = \lambda_F(x) \otimes S_{UD}(x) \bmod x^{[d-1]}$. We will show that $\psi_\ell = 0$ for $\ell = \tau,\ldots,d-2$.
%
%Let $\lambda_{FU}(x) = \lambda_F(x) \otimes \lambda_U(x)$ and $S_{FU}(x) = \lambda_{FU}(x) \otimes \bar{S}(x)$. According to (\ref{eq:lin-poly.product.small-left}), for $\epsilon + \mu \leq \ell \leq d-2$ we have
%\begin{align}
%  S_{FU,\ell} &= \sum_{i=0}^{\epsilon+\mu} \lambda_{FU,i} \bar{S}_{\ell-i}^{[i]} = \sum_{i=0}^{\epsilon+\mu}
%\lambda_{FU,i} \left(\sum_{j=1}^\tau E_j^{[\ell-i-d+2]} X_j \right)^{[i]} \nonumber \\
%  \label{eq:roots-lambda-D}
%  &= \sum_{j=1}^\tau E_j^{[\ell-d+2]} \lambda_{FU}(X_j) = \sum_{j=1}^{\delta} E_{\mu+j}^{[\ell-d+2]} \gamma_j,
%\end{align}
%where $\gamma_j = \lambda_{FU}(X_{\mu+j})$, $j=1,\ldots,\delta$.
%
%Note that $\lambda_F(x) \otimes S_{UD}(x) = S_{FU}(x) \otimes {\bar{\sigma}_D(x^{[d-2]})}^{[-d+2]}$. Using (\ref{eq:lin-poly.product.small-right}) and (\ref{eq:roots-lambda-D}), for $\delta+ \epsilon + \mu \leq \ell \leq d-2$ we have
%\begin{align}
%  \psi_\ell &= \sum_{i=0}^{\delta} {(\bar{\sigma}_{D,i}^{[-d+2]})}^{[\ell-i]} S_{FU,\ell-i} \nonumber \\
%  &= \sum_{i=0}^{\delta} \sigma_{D,\delta-i}^{[\ell-\delta-d+2]} \sum_{j=1}^{\delta} E_{\mu+j}^{[\ell-i-d+2]} \gamma_j \nonumber \\
%  &= \sum_{j=1}^{\delta} \sum_{i=0}^{\delta} \sigma_{D,i}^{[\ell-\delta-d+2]} E_{\mu+j}^{[\ell-\delta+i-d+2]} \gamma_j \nonumber \\
%  &= \sum_{j=1}^{\delta} \sigma_{D}(E_{\mu+j})^{[\ell-\delta-d+2]} \gamma_j = 0. \nonumber
%\end{align}
%This completes the proof of the theorem.
\end{IEEEproof}

The complete decoding algorithm based on the error locator polynomial
is given in Fig.~\ref{fig:dec-alg2}.
\begin{figure}
\hrulefill
\begin{center}
\begin{minipage}{1\columnwidth}
  \textbf{Input:} received tuple $(\bm{r}, \hat{L}, \hat{E}) \in \Fqm^n \times \Fq^{n \times \mu} \times \Fqm^\delta$. \\
  \textbf{Output:} error word $\bm{e} \in \Fqm^n$.
  \begin{enumerate}
  \item \emph{Computing the auxiliary syndrome polynomial}:\\ Compute
  \begin{enumerate}
    \item $S_\ell = \sum_{i=1}^n h_i^{[\ell]} r_i$, $\ell = 0,\ldots,d-2$
    \item $\hat{X}_j = \sum_{i=1}^{n} \hat{L}_{ij} h_i$, $j=1,\ldots,\mu$
    \item $\lambda_U(x) = \minpoly\{\hat{X}_1,\ldots,\hat{X}_\mu\}$
    \item $\sigma_D(x) = \minpoly\{\hat{E}_1,\ldots,\hat{E}_\delta\}$, and
    \item $S_{UD}(x) = \lambda_U(x) \otimes \bar{S}(x) \otimes \bar{\sigma}_D(x^{[d-2]})^{[-d+2]}$.
  \end{enumerate}
  \item \emph{Computing the error locator polynomial}:
  \begin{enumerate}
    \item Use the Berlekamp-Massey algorithm \cite{Richter.Plass2004:BerlekampMassey} to find $\lambda_F(x)$ that solve the key equation (\ref{eq:key-equation2}).
    \item Compute $S_{FU}(x) = \lambda_{F}(x) \otimes \lambda_{U}(x) \otimes \bar{S}(x)$.
    \item Use Gabidulin's algorithm \cite{Gabidulin1985} to find $\gamma_1,\ldots,\gamma_\delta \in \Fqm$ that solve \[
          S_{FU,\ell} = \sum_{j=1}^{\delta} E_{\mu+j}^{[\ell-d+2]} \gamma_j.
        \]
    \item Compute $\lambda_D(x) = \minpoly\{\gamma_1,\ldots,\gamma_\delta\}$ and
    \item $\lambda(x) = \lambda_D(x) \otimes \lambda_{F}(x) \otimes \lambda_{U}(x)$.
  \end{enumerate}
  \item \emph{Finding the roots of the error locator polynomial}:\\
     Use either the algorithm in \cite{Skachek.Roth2008} or the methods in \cite{Berlekamp} to find a basis $X_1,\ldots,X_\tau \in \Fqm$ for the root space of $\lambda(x)$.
  \item \emph{Finding the error values}:
  \begin{enumerate}
    \item Solve (\ref{eq:syndrome-equation}) using Gabidulin's algorithm \cite{Gabidulin1985} to find the error values $E_1,\ldots,E_\tau \in \Fqm$.
    \item Compute the error locations $L_{ij} = \sum_{k=1}^{m} X_{jk} Q_{ki}$, $i=1,\ldots,n$, $j=1,\ldots,\tau$.
    \item Compute the error word $\bm{e} = \sum_{j=1}^\tau L_j E_j$.
  \end{enumerate}
  \end{enumerate}
\end{minipage}
\end{center}
\hrulefill
  \caption{Generalized decoding algorithm for Gabidulin codes, alternative formulation.} \label{fig:dec-alg2}
\end{figure}

\subsection{Practical Considerations}

We have seen that the complexity of decoding a Gabidulin code $\mathcal{C} \subseteq \Fq^{n \times m}$ with $\dr(\mathcal{C}) = d$ is given by $O(dm)$ operations in $\Fqm$. In many applications, in particular for network coding, we have $m \gg n$. In such cases, the decoding complexity can be significantly reduced by using, rather than a Gabidulin code, an MRD code formed by the Cartesian product of many shorter Gabidulin codes with the same distance. More precisely, let $\ell = \lfloor \frac{m}{n} \rfloor$ and $n' = m - n(\ell-1)$. Take $\mathcal{C} = C_1 \times C_2 \times \cdots \times C_\ell$, where $C_i \subseteq \Fq^{n \times n}$, $i = 1,\ldots,\ell-1$, and $C_\ell \subseteq \Fq^{n \times n'}$ are Gabidulin codes with minimum rank distance $d$. Then $\mathcal{C}$ is an MRD code with $\dr(\mathcal{C}) = d$.

Now, decoding of $\mathcal{C}$ can be performed by decoding each $\mathcal{C}_i$ individually. Thus, assuming for simplicity that $m = n\ell$, the overall decoding complexity is given by $\ell O(dn) = O(dm)$ operations in $\mathbb{F}_{q^n}$. In other words, operations in a potentially large field $\Fqm$ can be replaced by operations in a much smaller field $\mathbb{F}_{q^n}$.

Note that, in this case, additional computational savings may be obtained, since all received words will share the same set of error locations. For instance, if all error locations are known and the decoding algorithm of Fig. \ref{fig:dec-alg2} is used, then only steps 1a), 1b) and 4a)--4c) need to be performed.

\section{Conclusions}
\label{sec:conclusions}

In this paper, we have introduced a new approach to the problem of
error control in random network coding. Our approach is based,
on the one hand, on K{\"o}tter and Kschischang's abstraction of the
problem as a coding-theoretic problem for subspaces and, on the other hand,
on the existence of optimal and efficiently-decodable codes for
the rank metric. We have shown that, when \emph{lifting} is performed
at the transmitter and \emph{reduction} at the receiver,
the random network coding channel behaves essentially as a matrix channel
that introduces errors in the rank metric and may also supply partial
information about these errors in the form of erasures and deviations.

An important consequence of our results is that many of the tools developed
for rank-metric codes can be \emph{almost} directly applied to random network
coding. However, in order to fully exploit the correction capability of
a rank-metric code, erasures and deviations must be taken into account. A second contribution of this work is the
generalization of the decoding algorithm for Gabidulin codes in order to
fulfill this task. Our proposed algorithm requires $O(dm)$ operations
in $\Fqm$, achieving the same complexity as conventional decoding algorithms that only correct rank errors.

Following this work, a natural step toward practical error control in
random network coding is the pursuit of efficient software
(and possibly hardware) implementations of encoders and decoders for Gabidulin codes.
Another avenue would be the investigation of more general network
coding scenarios where error and erasure correction might be useful;
for example, the case of multiple heterogeneous receivers can be
addressed using a priority encoding transmission scheme based
on Gabidulin codes \cite{Silva.Kschischang2007:PET}. An exciting open
question, paralleling the development of Reed-Solomon codes, is whether
an efficient list-decoder for Gabidulin codes exists that would allow
correction of errors above the error-correction bound.

We believe that, with respect to forward error (and erasure) correction,
Gabidulin codes will play the same role in random network coding that
Reed-Solomon codes have played in traditional communication systems.

\appendix

\subsection{Proof of Proposition~\ref{prop:reduction.existence}}

Before proving Proposition~\ref{prop:reduction.existence}, let us recall some properties of the matrices $I_\mathcal{U}$ and $I_{\mathcal{U}^c}$, where $I = I_{n \times n}$, $\mathcal{U} \subseteq \{1,\ldots,n\}$ and $\mathcal{U}^c = \{1,\ldots,n\} \setminus \mathcal{U}$.

For any $A \in \Fq^{n \times k}$ (respectively, $A \in \Fq^{k \times n}$), the matrix $I_{\mathcal{U}}^T A$ (resp., $A I_{\mathcal{U}}$) extracts the rows (resp., columns) of $A$ that are indexed by $\mathcal{U}$. Conversely, for any $B \in \Fq^{|\mathcal{U}| \times k}$ (resp., $B \in \Fq^{k \times |\mathcal{U}|}$) the matrix $I_{\mathcal{U}} B$ (resp., $B I_{\mathcal{U}}^T$) reallocates the rows (resp., columns) of $B$ to the positions indexed by $\mathcal{U}$, where all-zero rows (resp., columns) are inserted at the positions indexed by $\mathcal{U}^c$. Furthermore, observe that $I_\mathcal{U}$ and $I_{\mathcal{U}^c}$ satisfy the following properties:
\begin{align}
  I &= I_{\mathcal{U}}I_{\mathcal{U}}^T + I_{\mathcal{U}^c}I_{\mathcal{U}^c}^T, \nonumber \\
  I_{\mathcal{U}}^T I_{\mathcal{U}} &= I_{|\mathcal{U}| \times |\mathcal{U}|} \nonumber \\
  I_{\mathcal{U}}^T I_{\mathcal{U}^c} &= 0. \nonumber
\end{align}

We now give a proof of Proposition~\ref{prop:reduction.existence}.

\medskip
\begin{IEEEproof}[Proof of Proposition~\ref{prop:reduction.existence}]
Let $\RRE(Y)$ denote the reduced row echelon form of $Y$. For $i=1,\ldots,N$, let $p_i$ be the column position of the leading entry of row $i$ in $\RRE(Y)$. Let $\mathcal{U}^c = \{p_1,\ldots,p_{n-\mu}\}$ and $\mathcal{U} = \{1,\ldots,n\} \setminus \mathcal{U}^c$. Note that $|\mathcal{U}| = \mu$. From the properties of the reduced row echelon form, we can write
\begin{equation}\nonumber
  \RRE(Y) = \mat{W & \tilde{\bm{r}} \\ 0 & \hat{E}}
\end{equation}
where $\tilde{\bm{r}} \in \Fq^{(n-\mu) \times m}$, $\hat{E} \in \Fq^{\delta \times m}$ has rank $\delta$, and $W \in \Fq^{(n-\mu) \times n}$ satisfies $W I_{\mathcal{U}^c} = I_{(n-\mu) \times (n-\mu)}$.

Now, let
\begin{equation}\nonumber
  \bar{Y} = \mat{I_{\mathcal{U}^c} & 0 \\ 0 & I_{\delta \times \delta}} \RRE(Y) = \mat{I_{\mathcal{U}^c}W & \bm{r} \\ 0 & \hat{E}}
\end{equation}
where $\bm{r} = I_{\mathcal{U}^c}\tilde{\bm{r}}$. Since $I = I_{\mathcal{U}^c} I_{\mathcal{U}^c}^T + I_{\mathcal{U}} I_{\mathcal{U}}^T$, we have
\begin{align}
  I_{\mathcal{U}^c}W &= I_{\mathcal{U}^c} W (I_{\mathcal{U}^c} I_{\mathcal{U}^c}^T + I_{\mathcal{U}} I_{\mathcal{U}}^T) \nonumber \\
  &= I_{\mathcal{U}^c} I_{\mathcal{U}^c}^T + I_{\mathcal{U}^c} W I_{\mathcal{U}} I_{\mathcal{U}}^T \nonumber \\
  &= I - I_{\mathcal{U}} I_{\mathcal{U}}^T + I_{\mathcal{U}^c} W I_{\mathcal{U}} I_{\mathcal{U}}^T \nonumber \\
  &= I + \hat{L} I_{\mathcal{U}}^T \nonumber
\end{align}
where $\hat{L} = -I_{\mathcal{U}} + I_{\mathcal{U}^c} W I_{\mathcal{U}}$.
Also, since $I_{\mathcal{U}}^T I_{\mathcal{U}} = I_{\mu \times \mu}$ and $I_{\mathcal{U}}^T I_{\mathcal{U}^c} = 0$, we have $I_{\mathcal{U}}^T \hat{L} = - I_{\mu \times \mu}$ and $I_{\mathcal{U}}^T \bm{r} = 0$.

Thus,
\begin{equation}\nonumber
  \bar{Y} = \mat{I + \hat{L} I_{\mathcal{U}}^T & \bm{r} \\ 0 & \hat{E}}
\end{equation}
is a matrix with the same row space as $Y$. The proof is complete.
\end{IEEEproof}
%\medskip

\subsection{Proof of Proposition~\ref{prop:reduction.redundant-set}}

\begin{IEEEproof}[Proof of Proposition~\ref{prop:reduction.redundant-set}]
  We want to show that
  \begin{equation}\nonumber
    \linspan{\mat{I + \hat{L}TI_{\mathcal{S}}^T & \bm{r} \\ 0 & R\hat{E}}} = \linspan{\mat{I + \hat{L}I_{\mathcal{U}}^T & \bm{r} \\ 0 & \hat{E}}}.
  \end{equation}
  From (\ref{eq:stacked-matrices-subspace-sum}) and the fact that $R$ is nonsingular (since $\rank R\hat{E} = \delta$), this amounts to showing that
  \begin{equation}\nonumber
    \linspan{\mat{I + \hat{L}TI_{\mathcal{S}}^T & \bm{r}}} = \linspan{\mat{I + \hat{L}I_{\mathcal{U}}^T & \bm{r}}}.
  \end{equation}
  Let $W_1 = I + \hat{L}I_{\mathcal{U}}^T$ and $W_2 = I + \hat{L}TI_{\mathcal{S}}^T$. Note that, since $W_1 I_{\mathcal{U}^c} = I_{\mathcal{U}^c}$ and $I_\mathcal{U}^T \mat{W_1 & \bm{r}} = 0$, we have that $I_{\mathcal{U}^c}^T W_1$ is full rank. Similarly, $I_{\mathcal{S}}^T \mat{W_2 & \bm{r}} = 0$ and $I_{\mathcal{S}^c}^T W_2$ is full rank. Thus, it suffices to prove that
  \begin{equation}\label{eq:proof-redundant-1}
    M \mat{W_2 & \bm{r}} = \mat{W_1 & \bm{r}}
  \end{equation}
  for some $M \in \Fq^{n \times n}$.

  Let $\mathcal{A}  = \mathcal{U} \cup \mathcal{S}$ and $\mathcal{B}  = \mathcal{U} \cap \mathcal{S}$. Observe that $M$ can be partitioned into three sub-matrices, $M I_{\mathcal{A}^c}$, $M I_{\mathcal{S}}$ and $M I_{\mathcal{U}\setminus \mathcal{B}}$. Choose $M I_{\mathcal{A}^c} = I_{\mathcal{A}^c}$, and $M I_{\mathcal{S}}$ arbitrarily. We will choose $M I_{\mathcal{U}\setminus \mathcal{B}}$ so that (\ref{eq:proof-redundant-1}) is satisfied. First, note that
  \begin{equation}\nonumber
    M\bm{r} = M(I_{\mathcal{A}^c}I_{\mathcal{A}^c}^T + I_{\mathcal{A}}I_{\mathcal{A}}^T)\bm{r} = I_{\mathcal{A}^c}I_{\mathcal{A}^c}^T \bm{r} = \bm{r}
  \end{equation}
  since $I_{\mathcal{A}}^T \bm{r} = 0$. Thus, we just need to consider $M W_2 = W_1$ in (\ref{eq:proof-redundant-1}). Moreover, note that
  \begin{align}
  M W_2 &= M(I_{\mathcal{A}^c}I_{\mathcal{A}^c}^T + I_{\mathcal{S}}I_{\mathcal{S}}^T + I_{\mathcal{U}\setminus\mathcal{B}}I_{\mathcal{U}\setminus\mathcal{B}}^T) W_2 \nonumber \\
  &= I_{\mathcal{A}^c}I_{\mathcal{A}^c}^T W_2 + (M I_{\mathcal{U}\setminus\mathcal{B}}) (I_{\mathcal{U}\setminus\mathcal{B}}^T W_2). \nonumber
  \end{align}
  Now, consider the system $M W_2 = W_1$. From basic linear algebra, we can solve for $M I_{\mathcal{U}\setminus\mathcal{B}}$ if and only if
  \begin{equation}\nonumber
    \rank \mat{I_{\mathcal{U}\setminus\mathcal{B}}^T W_2 \\ W_1 - I_{\mathcal{A}^c}I_{\mathcal{A}^c}^T W_2} \leq |\mathcal{U}\setminus\mathcal{B}|.
  \end{equation}
  Since $I_{\mathcal{U}\setminus\mathcal{B}}^T W_1 = 0$ and $I_{\mathcal{S}}^T W_2 = 0$, we can rearrange rows to obtain
\begin{align}
  \rank \!\mat{I_{\mathcal{U}\setminus\mathcal{B}}^T W_2 \\ W_1 - I_{\mathcal{A}^c}I_{\mathcal{A}^c}^T W_2}
 &= \rank \!\mat{I_{\mathcal{U}\setminus\mathcal{B}}^T (W_1 - W_2) \\ I_{(\mathcal{U}\setminus\mathcal{B})^c}I_{(\mathcal{U}\setminus\mathcal{B})^c}^T (W_1 - W_2)} \nonumber \\
 &= \rank (W_1 - W_2). \nonumber
\end{align}
  To complete the proof, we will show that $\rank(W_1 - W_2) \leq |\mathcal{U}\setminus \mathcal{B}|$. We have
  \begin{align}
  \rank(W_1 - W_2)
  &= \rank(\hat{L}I_\mathcal{U}^T - \hat{L}T I_\mathcal{S}^T) \nonumber \\
%  &= \rank \hat{L}(I_\mathcal{U}^T - T I_\mathcal{S}^T) \nonumber \\
  &\leq \rank (I_\mathcal{U}^T - T I_\mathcal{S}^T) \nonumber \\
  &= \rank (I_\mathcal{S}^T \hat{L} I_\mathcal{U}^T + I_\mathcal{S}^T) \label{eq:proof-redundant-5} \\
  &= \rank I_\mathcal{S}^T W_1 \nonumber \\
  &= \rank I_\mathcal{S} I_\mathcal{S}^T W_1 \nonumber \\
  &= \rank (I_{\mathcal{S}\setminus\mathcal{B}} I_{\mathcal{S}\setminus\mathcal{B}}^T + I_\mathcal{B} I_\mathcal{B}^T) W_1 \nonumber \\
  &= \rank I_{\mathcal{S}\setminus\mathcal{B}} I_{\mathcal{S}\setminus\mathcal{B}}^T W_1 \nonumber \\
  &\leq |\mathcal{S}\setminus\mathcal{B}| = |\mathcal{U}\setminus\mathcal{B}|. \nonumber
  \end{align}
  where (\ref{eq:proof-redundant-5}) is obtained by left multiplying by $I_\mathcal{S}^T \hat{L} = -T^{-1}$.
\end{IEEEproof}
%\medskip

\subsection{Proof of Theorem~\ref{thm:reduction-and-Delta}}

\begin{IEEEproof}[Proof of Theorem~\ref{thm:reduction-and-Delta}]
We have
\begin{align}
\rank \mat{X \\ Y}
&= \rank \mat{I & \bm{x} \\ I + \hat{L}I_\mathcal{U}^T & \bm{r} \\ 0 & \hat{E}} \nonumber \\
&= \rank \mat{- \hat{L}I_\mathcal{U}^T & \bm{x}-\bm{r} \\ I + \hat{L}I_\mathcal{U}^T & \bm{r} \\ 0 & \hat{E}} \nonumber \\
&= \rank \mat{\hat{L}I_\mathcal{U}^T & \bm{r}-\bm{x} \\ I_{\mathcal{U}^c}^T (I + \hat{L}I_\mathcal{U}^T) & I_{\mathcal{U}^c}^T \bm{r} \\ 0 & \hat{E}} \label{eq:proof-reduc-3} \\
&= \rank \mat{\hat{L}I_\mathcal{U}^T & \bm{r}-\bm{x} \\ I_{\mathcal{U}^c}^T & I_{\mathcal{U}^c}^T \bm{x} \\ 0 & \hat{E}} \label{eq:proof-reduc-4} \\
&= \rank \mat{\hat{L}I_\mathcal{U}^T & \bm{r}-\bm{x} \\ 0 & \hat{E}} + \rank \mat{I_{\mathcal{U}^c}^T & I_{\mathcal{U}^c}^T \bm{x}} \label{eq:proof-reduc-5} \\
&= \rank \mat{\hat{L} & \bm{r}-\bm{x} \\ 0 & \hat{E}} + n-\mu \label{eq:proof-reduc-6}
\end{align}
where (\ref{eq:proof-reduc-3}) follows from $I_{\mathcal{U}}^T \mat{I + \hat{L}I_\mathcal{U}^T & \bm{r}} = 0$, (\ref{eq:proof-reduc-5}) follows by subtracting $I_{\mathcal{U}^c}^T \mat{\hat{L}I_\mathcal{U}^T & \bm{r}-\bm{x}}$ from $\mat{I_{\mathcal{U}^c}^T (I + \hat{L}I_\mathcal{U}^T) & I_{\mathcal{U}^c}^T \bm{r}}$, (\ref{eq:proof-reduc-5}) follows from $I_{\mathcal{U}^c}^T I_{\mathcal{U}^c} = I_{(n-\mu) \times (n-\mu)}$ and $\hat{L}I_\mathcal{U}^T I_{\mathcal{U}^c} = 0$ (i.e., the two matrices in (\ref{eq:proof-reduc-5}) have row spaces that intersect trivially), and (\ref{eq:proof-reduc-6}) follows by deleting the all-zero columns.

Since $\rank X + \rank Y = 2n -\mu + \delta$, we have
\begin{align}
d_S(\linspan{X},\linspan{Y}) &= 2 \rank \mat{X \\ Y} - \rank X - \rank Y \nonumber \\
&= 2\rank \mat{\hat{L} & \bm{r}-\bm{x} \\ 0 & \hat{E}} - \mu - \delta.  \nonumber
\end{align}
\end{IEEEproof}
%\medskip

\subsection{Proof of Proposition~\ref{prop:rank-expansion.equivalence}}

Before proving Proposition~\ref{prop:rank-expansion.equivalence}, we need the following lemma.

\medskip
\begin{lemma} \label{lem:min-rank.unconstrained}
  For $X \in \Fq^{n \times m}$ and $Y \in \Fq^{N \times m}$ we have
  \begin{equation}\nonumber
    \min_{A \in \Fq^{N \times n}} \rank(Y - AX) = \rank \mat{X \\ Y} - \rank X
  \end{equation}
  and for $X \in \Fq^{n \times m}$ and $Y \in \Fq^{n \times M}$ we have
  \begin{equation}\nonumber
    \min_{B \in \Fq^{m \times M}} \rank(Y - XB) = \rank \mat{X & Y} - \rank X.
  \end{equation}
\end{lemma}
\begin{IEEEproof}
  For any $A \in \Fq^{N \times n}$, we have
%  \begin{align}
%  \rank \mat{X \\ Y} &= \rank \mat{X \\ Y - AX} \nonumber \\
%  &\leq \rank X + \rank (Y-AX) \nonumber
%  \end{align}
  \begin{equation} \nonumber
    \rank \mat{X \\ Y} = \rank \mat{X \\ Y - AX} \leq \rank X + \rank (Y-AX)
  \end{equation}
  which gives a lower bound on $\rank (Y - AX)$. We now prove that this lower bound is achievable.

  Let $Z \in \Fq^{t \times m}$ be such that $\linspan{Y} = \linspan{X} \cap \linspan{Y} \oplus \linspan{Z}$, where $t = \rank Y - \omega$ and $\omega = \dim \linspan{X} \cap \linspan{Y}$. Let $B \in \Fq^{\omega \times n}$ be such that $\linspan{BX} = \linspan{X} \cap \linspan{Y}$. We can write $Y = T \mat{BX \\ Z}$ for some full-rank $T \in \Fq^{N \times (\omega + t)}$. Now, let $A = T \mat{B \\ 0} \in \Fq^{N \times m}$. Then
  \begin{align}
  \rank (Y - AX)
  &= \rank(T \mat{BX \\ Z} - T \mat{BX \\ 0}) \nonumber \\
  &= \rank(T \mat{0 \\ Z}) = \rank Z \nonumber \\
  &= \rank Y - \dim (\linspan{X} \cap \linspan{Y}) \nonumber \\
  &= \rank \mat{X \\ Y} - \rank X. \nonumber
  \end{align}
  This proves the first statement. The second statement is just the transposed version of the first one.
\end{IEEEproof}
\medskip

\begin{IEEEproof}[Proof of Proposition~\ref{prop:rank-expansion.equivalence}]
Let
\begin{equation}\nonumber
  \epsilon' = \min_{E^{(1)}, L^{(2)}}\, \rank (\bm{e} - \hat{L} E^{(1)} - L^{(2)} \hat{E}).
\end{equation}

We first show the equivalence of 1) and 2). From Lemma~\ref{lem:min-rank.unconstrained}, we have
\begin{equation}\nonumber
  \min_{L^{(2)}}\, \rank (e - \hat{L} E^{(1)} - L^{(2)}\hat{E}) = \rank \mat{e - \hat{L} E^{(1)} \\ \hat{E}} - \rank \hat{E}.
\end{equation}
Similarly, from Lemma~\ref{lem:min-rank.unconstrained} we have
\begin{align}
\min_{E^{(1)}}\, \rank \mat{e - \hat{L} E^{(1)} \\ \hat{E}}
&= \min_{E^{(1)}}\, \rank \left( \mat{e \\ \hat{E}} - \mat{\hat{L} \\ 0} E^{(1)} \right) \nonumber \\
&= \rank \mat{\hat{L} & \bm{e} \\ 0 & \hat{E}} - \rank \hat{L}. \nonumber
\end{align}
Thus,
\begin{equation}\nonumber
  \epsilon' = \rank \mat{\hat{L} & \bm{e} \\ 0 & \hat{E}} - \mu - \delta
\end{equation}
and the equivalence is shown.

Now, observe that the statement in 3) is equivalent to the statement that $\tau^* - \mu - \delta$ is the minimum value of $\epsilon$ for which there exist $E^{(1)} \in \Fq^{\mu \times m}$, $L^{(2)} \in \Fq^{n \times \delta}$, $L^{(3)} \in \Fq^{n \times \epsilon}$ and $E^{(3)} \in \Fq^{\epsilon \times m}$ satisfying
\begin{equation}\nonumber
  \bm{e} = \hat{L} E^{(1)} + L^{(2)} \hat{E} + L^{(3)} E^{(3)}.
\end{equation}
To show the equivalence of 2) and 3), we will show that $\epsilon' = \epsilon''$, where
\begin{align}
\epsilon'' &= \min_{\substack{\epsilon, E^{(1)}, L^{(2)}, L^{(3)}, E^{(3)}: \\ \bm{e} = \hat{L} E^{(1)} + L^{(2)} \hat{E} + L^{(3)} E^{(3)}}}\, \epsilon. \nonumber
\end{align}
We can rewrite $\epsilon''$ as
\begin{align}
\epsilon''
&= \min_{E^{(1)}, L^{(2)}}\, \min_{\substack{\epsilon, L^{(3)}, E^{(3)}: \\ \bm{e} - \hat{L} E^{(1)} - L^{(2)} \hat{E} = L^{(3)} E^{(3)}}}\, \epsilon \nonumber \\
&= \min_{E^{(1)}, L^{(2)}}\, \rank (\bm{e} - \hat{L} E^{(1)} - L^{(2)} \hat{E}) \label{eq:proof-equiv-1} \\
&= \epsilon'. \nonumber
\end{align}
where (\ref{eq:proof-equiv-1}) follows from (\ref{eq:rank.equivalent-definition}). This shows the equivalence between 2) and 3).
\end{IEEEproof}

\end{document}